\documentstyle[11pt]{article}
\newcommand{\be}{\begin{equation}}
\newcommand{\ee}{\end{equation}}
\newcommand{\ba}{\begin{eqnarray}}
\newcommand{\ea}{\end{eqnarray}}
\newcommand{\ket}[1]{| {#1} \rangle}
\newcommand{\bra}[1]{\langle {#1} |}
\newcommand{\ave}[1]{\langle {#1} \rangle}

\begin{document}
\begin{titlepage}
\vspace*{3.5cm}
\begin{center}
\begin{Large}
{\bf The Linear Sigma-Model in the $1/N$-Expansion via Dynamical\\
Boson Mappings and Applications to $\pi\pi$-Scattering}
\end{Large}
\vspace*{1.cm}

Z. Aouissat$^{(1)}$,  P. Schuck$^{(2)}$ and J. Wambach$^{(1)}$,\\
$^{(1)}$ {\small{\it Institut f\"{u}r Kernphysik, Technische Hochschule 
Darmstadt, Schlo{\ss}gartenstra{\ss}e 9,\\ D-64289 Darmstadt, Germany.}}\\
$^{(2)}$ {\small{\it CNRS-IN2P3 Universit\'{e} Joseph Fourier,
 Institut des Sciences Nucl\'{e}aires, 53 avenue des Martyrs,\\
F-38026 Grenoble C\'edex, France.}}\\
\end{center}
\vspace*{2.cm}
\begin{abstract}
We present a non-perturbative method for the study of the $O(N+1)$-version
of the linear sigma-model. Using boson-mapping techniques, in close analogy 
to those well-known for fermionic systems, we obtain a systematic $1/N$-expansion
for the Hamiltonian which is symmetry-conserving order by order. 
The leading order for the Hamiltonian is evaluated explicitly and we apply the
method to $\pi\pi$-scattering, in deriving the $T$-matrix to leading 
order.
\end{abstract}

\end{titlepage}
\newpage

\section{Introduction}
In a recent publication \cite{ACSW} we have presented a non-perturbative 
treatment of the pion as a Goldstone particle  within the linear sigma-model 
using the Hartree-Fock-Bogoliubov (HFB)-RPA theory. 
Our initial interest in a non-perturbative approach was motivated by the 
desire to find a pion-pion scattering equation 
which obeys unitarity while being consistent with constraints from 
chiral symmetry, such as the vanishing of the $a_0^0$ scattering length in 
the chiral limit, $m_\pi\to 0$. This is nontrivial since, even in cases
where the tree-level scattering amplitude fulfills the chiral symmetry 
requirements, the iteration through various reduction schemes of the 
Bethe-Salpeter equation, such as time-ordered perturbation theory or the 
Blankenbecler-Sugar reduction breaks the consistency \cite{ARCSW}.
 Unitary scattering 
equations are a necessity, however, when studying medium modifications of
elementary cross sections. Indeed for example a $\pi^+$-$\pi^-$ 
pair in a hot pion gas may form a resonance close to the
$2m_{\pi}$ threshold or even a bound state as was discussed in 
\cite{ACSW92, ACSW96, RW}. Of course such features cannot be treated in 
chiral
perturbation theory. In the present work we will therefore generalize the 
theory to the two-pion problem with application to $\pi$-$\pi$ scattering 
and will indeed find a scattering equation which fulfills the requirements
of chiral symmetry.\\
In constructing the physical pion we have used in our previous work 
\cite{ACSW} a RPA operator which mixes a quasi-pion with a 
quasi-pion quasi-sigma pair, obtained from a HFB mean-field calculation, in 
such a way that this operator contains the axial charges $Q^i_5$ as a 
special limit. The resulting RPA-eigenmode, at zero three-momentum, is then 
 identified with the physical pion and we were able to show that it 
becomes a Goldstone mode in the chiral limit. To fully restore chiral 
symmetry, which is broken at the HFB level, it is important to write down 
the RPA operator in the selfconsistent HFB basis whose vacuum is 
characterized by the coherent (squeezed) state 
$\ket{ \Phi} \propto \exp\left[ (\sum_q z_1(q)\vec a_q^+\cdot\vec a_{-q}^+ 
\, + \, z_2(q) b_q^+b_{-q}^+)\, +\,  w b_0^+ \right] \ket{0}$. 
The z's and $ w$ are variational parameters and $a^+$, $b^+$ are the original 
pion and sigma creation operators corresponding to the vacuum defined by
$a^i_q\ket{0}= b_q\ket{0}=0$.
Although this approach is fairly standard and described in most textbooks 
(see e.g. \cite{BLRI, RS80}) the field theoretical context exhibits some 
particularities which are worthwhile mentioning. The main point is that the
selfconsistent HFB basis generates a mass for the quasipion even in the chiral
limit. Despite the fact that they emerge from a respectable Raleigh-Ritz 
variational principle and that Goldstone's theorem is, as it should be, 
perfectly restored once the RPA fluctuations are included this remains a 
somewhat disturbing feature. In particular for the two-pion problem 
a serious problem emerges since the asymptotic states cannot be 
defined easily in a way which is consistent which chiral symmetry. The 
HFB quasiparticles can clearly not be used. Even though the two-pion
RPA, when solved in a similar fashion as the single-pion RPA, will exhibit
a spurious mode in the chiral limit, this does not imply that the 
scattering length will vanish. Interactions will persist so as to cancel 
the quasiparticle mass term, in close analogy to the single-pion case. This is
obviously unphysical. In addition, with explicit symmetry breaking, the 
threshold will be at twice the quasipion mass and not at twice the 
physical mass. Because of the obvious advantages of a Raleigh-Ritz 
variational theory it maybe worthwhile for the future to consider
modifications of the trial wave function to guarantee that Goldstone bosons 
remain as such. In this paper we will not pursue this option despite 
the fact that the approach of \cite{ACSW} might probably be generalized to 
the two-pion problem. Instead we will identify the origin of the mass
generation in the HFB mean field by studying the $O(N+1)$ version of
the linear sigma-model. This will lead to a systematic  
$1/N$-expansion technique  which has the advantage of being
non-perturbative as well (it will also lead to RPA type equations).
It will be shown that, to leading order, the pion is a Goldstone boson
opening the possibility to also derive a $\pi-\pi$ scattering
equation to leading order which is physically meaningful. The 
disadvantage is that the Raleigh-Ritz variational principle 
is lost. In regards to earlier work on non-perturbative approaches to the $\Phi^4$ field theories in general \cite{Coop, Kerm} and  
the $1/N$-expansion in particular  \cite{BG, Abot, Root} we here will however 
present a novel approach. It employs Boson expansion methods
and relies on work in nuclear physics where such techniques, in 
connection with the restoration of spontaneously broken symmetries, have been 
studied extensively \cite{M74, M93}. Boson expansion 
methods for boson pairs in interacting Bose systems have first been 
introduced by Curutchet, Dukelsky, Dussel, and Fendrik \cite{CDDF} and further elaborated by  Bijker, Pittel and Dukelsky in \cite{PIDUK}, in close analogy 
to the techniques known from the fermion case \cite{BLRI, RS80}.\\
Since the subject of restoration of broken symmetries is full of subtleties, 
it seems worthwhile to present this approach in some detail. Our
aim will be to derive a covariant scattering equation for two 
pions to-leading order without taking the 
large-$N$ limit. This will be achieved by a systematic $1/N$-expansion of 
the Hamiltonian of the linear sigma-model. Of course in reality $N=3$ and 
it will be necessary to work out corrections to the lowest order $\pi-\pi$ 
scattering equation for quantitative studies.
 Within the bosonization framework this can be done
systematically and we will leave the evaluation of the $1/N$ corrections
 for future work.\\
In detail our paper is organized as follows. First we recapitulate the
most important results of our previous work, here applied, however, to the 
$O(N+1)$-version of the linear sigma-model. As mentioned above, this 
mainly serves to elucidate the origin of the mass generation at the HFB 
level.  In sect.~3 the boson mapping techniques are introduced which lead
to the $1/N$ expansion of the sigma-model Hamiltonian. The leading-order 
results are then represented. For 
the two-pion sector the T-matrix will be extracted in sections ~4 and 5
 where we
also comment on the symmetry properties. Conclusions and an outlook will 
be given in sect.~6.

\section{HFB-RPA for the  $O(N+1)$ Sigma-Model}

As mentioned in the introduction this section serves to identify the
origin of the mass generation in the HFB-RPA approach of ref.~\cite{ACSW}
from the point of view of a $1/N$-expansion. We therefore
consider the $O(N+1)$ version of the linear sigma-model.
The difference to the $SU(2)\times SU(2)\sim O(4)$ case lies in the fact 
that one now has an $N$-component isovector pion field. The Lagrangian 
density then reads  
\\
\begin{equation}
 {\cal L}  =  \frac{1}{2}\left[ \left(\partial_{\mu}{\vec{\pi}}\right)^2
     + \left(\partial_{\mu}{\hat \sigma}\right)^2 \right]
 - \frac{\mu_{0}^2}{2} \left[ {\vec{\pi}}^2 + {\hat \sigma}^2 \right]
 - \frac{\lambda_{0}^2}{4N}\left[ {\vec{\pi}}^2 + {\hat \sigma}^2 \right]^2
   + \sqrt{N} c {\hat \sigma}
\label{eq201}
\end{equation}
\\
where $\lambda_0$ represents the bare coupling constant, $\mu_0$ the bare
mass parameter and $\pi$ and ${\hat \sigma}$ denote the bare pion and sigma
fields, respectively. Chiral symmetry is explicitly broken (in the PCAC
sense) by the last term in the Lagrangian, $c {\hat \sigma}$. \\

It is now convenient to define the field operators in terms of creation and 
annihilation operators as
\begin{eqnarray}
  \vec\pi({\bf x}) &=& \int \frac{d^3 q}
  {\sqrt{(2\pi)^{3} 2\omega_{q}}}
  \left(\vec a_qe^{i{\bf q}{\bf x}} +\vec a_q^+
  e^{-i{\bf q}{\bf x}} \right) \nonumber\\
  {\hat \sigma}({\bf x}) &=& \int \frac{d^3 q}
  {\sqrt{(2\pi)^{3} 2\omega_{q}}}
  \left(b_ q e^{i{\bf q}{\bf x}} + b^{+}_q
  e^{-i{\bf q}{\bf x}} \right)
\label{eq202}
\end{eqnarray}
where the frequency $\omega_q$, common to both fields, is given by
\begin{equation}
      \omega_q = \sqrt{\mu_0^2 + q^2}.
\label{eq203}
\end{equation}
Next a canonical transformation is performed for the pion- as 
well as the sigma- fields by introducing a new set of creation and 
annihilation operators through a Bogoliubov rotation 
\\
\begin{eqnarray}
  \vec\alpha_q^+ &=& u_{q}\vec a_q^+ - v_{q}\vec a_{-q},
  \nonumber\\
  \beta ^{+}_{q} &=& x_{q}b^{+}_{q} -
    y_{q}b_{-q} - w^*\,\delta_{q0}
\label{eq204}
\end{eqnarray}
\\
with $u_{q}$, $v_{q}$,  $x_{q}$ and $y_{q}$ being even functions of their 
argument, and $w$ a c-number. The additional 
'shift' in the second equation accounts for the macroscopic condensate
\\
\begin{equation}
\ave{\hat \sigma}\,=\,\frac{\ave{b_0^+}+\ave{b_0}}{\sqrt{(2\pi)^3 2\mu_0}}
\, =\, \frac{(x_0+y_0)(w\,+\,w^*)}{\sqrt{(2\pi)^3 2\mu_0}} \,=\, \sqrt{N} s
\label{resq1}
\end{equation}
\\
To render the transformations canonical the 
Bogoliubov factors have to obey the constraints
\\
\begin{equation}
       u_{q}^{2} -  v_{q}^{2} = 1, \quad\quad\quad
       x_{q}^{2} -  y_{q}^{2} =1.
\label{eq205}
\end{equation}
\\
The quasiparticle vacuum  $\ket{ \Phi}$ 
( $\vec \alpha  \ket{ \Phi}= \beta  \ket{ \Phi} = 0$ )
is now given by the following coherent state
\\
\begin{equation}
\ket{ \Phi} =\exp\left[ \sum_q ( z_1(q)\vec a_q^+\cdot\vec a_{-q}^+ 
\, + \, z_2(q) b_q^+
b_{-q}^+)\, +\, \frac{w}{x_0} b_0^+ \right] \ket{0}.
\label{eq206}
\end{equation}
\\
were $\ket{0}$ denotes the vacuum for the original basis 
($\vec a_q\ket{ 0 } =b_q \ket{ 0 } =0$ )
and  $ z_1 = \frac{v}{2\,u},\quad z_2= \frac{y}{2\,x}$.\\
It is  straightforward to write the Hamiltonian 
in the quasiparticle basis. The amplitudes $u,v,x,y$ as well as 
the condensate $s$ are determined by minimizing
the vacuum expectation value $\bra{\Phi } H \ket{ \Phi }/
\ave{\Phi\mid \Phi}$. This is equivalent to demanding that the 
single-particle part of $H$ is diagonal and leads to the well-known BCS 
gap equations and an equation which determines $s$. 
These can be cast in a form which contains physical quantities:   
\\
\begin{eqnarray}
{\cal E}_{\pi}^2&=& \mu_0^2 + 
\lambda_0^2 \left[\frac{N+2}{N}I_{\pi} +\frac{1}{N}I_{\sigma} + s^2 \right]
\nonumber\\
{\cal E}_{\sigma}^2&=& \mu_0^2 + \lambda_0^2 \left[I_{\pi} 
 +\frac{3}{N}I_{\sigma} + 3 s^2 \right]
\nonumber\\
\frac{c}{s} &=&  \mu_0^{2} + 
\lambda_0^{2} \left[ I_{\pi} + \frac{3}{N}J_{\sigma} + s^{2} \right] .
\label{eq207}
\end{eqnarray}
\\
where ${\cal E}_{\pi}$ and ${\cal E}_{\sigma}$ denote the masses of the
quasi-pion and quasi-sigma respectively and $I_\pi$ and $I_\sigma$ are 
loop integrals
\begin{equation}
  I_{\pi}  =  \int \frac{d^{4}q}{(2\pi)^{4}}
  \quad \frac{i}{q^2-{\cal E}_{\pi}^2 +i\eta}, \quad \quad \quad
  I_{\sigma} = \int \frac{d^{4}q}{(2\pi)^{4}}
  \quad \frac{i}{q^2-{\cal E}_{\sigma}^2 + i\eta}.
\label{208}
\end{equation}
\\
For $N=3$ these results coincide with those of ref.~\cite{ACSW}.
By rewriting the equations for the quasiparticle masses as 
\\
\begin{eqnarray}
 {\cal E}_{\pi}^2(0)&=& \frac{c}{s}
  + {2\lambda_0^2\over N} \left[I_\pi -I_\sigma \right],
 \nonumber\\
 {\cal E}_{\sigma}^2(0)&=& \frac{c}{s} +2\lambda_0^2 s^2.
\label{eq209}
\end{eqnarray}
\\
one sees that, in the chiral limit ($c\to 0$), the quasipion mass remains
massive due to the nonvanishing difference $ I_\pi -I_\sigma$. 
At first glance this seems to be a $1/N$ effect, however due to the 
selfconsistency one actually has a power series in $1/N$!\\ 
To render the pion massless one needs to go further and include a set of 
well-defined RPA fluctuations. 
The procedure is described in detail in ref.~\cite{ACSW}
and thus we will be very brief here. 
One introduces an excitation operator with the quantum 
numbers of the pion which is, at most, bilinear in the creation and 
destruction operators
\\
\begin{equation}
\vec Q_{\pi}^+ =
 X^1_{\pi} \vec \alpha^{+}_0 \,\, -\,\,  Y^1_{\pi} \vec \alpha_0
 \quad+\quad
 \sum_{q}
  \left[ X_{\pi}^2(q)  \beta^+_q \vec \alpha^{+}_{-q}
 \, \,-\,\,  Y_{\pi}^2(q)  \beta_{-q} \vec \alpha_q \right]
\label{eq210}
\end{equation}
\\
and similarly for the sigma field
\begin{equation}
 Q_{\sigma}^+  \,=\, \left[
 X^1_{\sigma} \beta^{+}_0 \,\, -\,\,  Y^1_{\sigma} \beta_0
 \right]
 \,\,+\,\, 
  \sum_{q}
  \left[ X_{\sigma}^2(q)  \beta^+_q \beta^+_{-q}
 \, \,-\,\,Y_{\sigma}^2(q)  \beta_{-q} \beta_q \right]
  \,+\, 
 \sum_{q} \left[
 X_{\sigma}^3(q)  \vec \alpha^{+}_q\cdot \vec \alpha^{+}_{-q}
 \, \,-\,\, Y_{\sigma}^3(q)  \vec \alpha_{-q}\cdot \vec\alpha_{q}
 \right] .
\label{eq211}
\end{equation}
\\ 
The RPA ground state is determined by $\vec Q_\pi\ket{RPA}=
Q_\sigma\ket{RPA}=0$. Within the equation of motion method \cite{RS80, ROW} the
RPA frequencies can now be determined 
which finally leads to the following equation for the physical pion mass 
\\
\begin{equation}
m_{\pi}^2 \,=\, \frac{c}{s} \quad+\quad  \frac{ 2\lambda_0^2}{N}\,
\frac{  \left[{\cal E}_{\pi}^2 \,-\,
 {\cal E}_{\sigma}^2 \right]
\left[\Sigma_{\pi\sigma}(0)
 \,-\, \Sigma_{\pi\sigma}(m_{\pi}^2)\right]}
{ 1\quad-\quad \frac{2 \lambda_0^2}{N} \Sigma_{\pi\sigma}(m_{\pi}^2)}
\label{eq212}
\end{equation}
\\
where $\Sigma_{\pi\sigma}(p^2)$ is the quasipion-quasisigma bubble given 
by
\\
\be
\Sigma_{\pi\sigma}(p^2) = -i\int \frac{d^{4}q}{(2\pi)^{4}}
\frac{1}{q^2-{\cal E}_{\pi}^2 +i\eta }\,\,
 \frac{1}{(p-q)^2-{\cal E}_{\sigma}^2 +i\eta} .
\label{eq213}
\ee
\\
In the chiral limit $(c=0)$ the Goldstone 
theorem is now manifest since eq.~(\ref{eq212}) has a zero-energy 
solution. Here again one sees that the RPA has brought about a power series 
in $1/N$ which, in the chiral limit, cancels exactly the mean field contributions
in such a way that the symmetry is restored.\\
After the inclusion of RPA fluctuations one obtains for the sigma mass   
\\
\begin{equation}
m^2_{\sigma}\,=\, {\cal E}^2_{\sigma} \quad+\quad
2 \lambda_0^4 s^2
\frac{ \Sigma_{\pi\pi}(m^2_{\sigma})\,\,+\,\,\frac{9}{N}\Sigma_{\sigma\sigma}(m^2_{\sigma})\,\,-\,\, 6\lambda_0^2 \frac{N+3}{N^2} \Sigma_{\pi\pi}(m^2_{\sigma})\Sigma_{\sigma\sigma}(m^2_{\sigma}) }
{\left[ 1\,-\, \frac{N+2}{N} \lambda_0^2 \Sigma_{\pi\pi}(m^2_{\sigma})\right]
\left[  1\,-\, \frac{3}{N} \lambda_0^2 \Sigma_{\sigma\sigma}(m^2_{\sigma})\right]
\,\,-\,\, \frac{1}{N} \lambda_{0}^4\Sigma_{\pi\pi}(m^2_{\sigma})\Sigma_{\sigma\sigma}(m^2_{\sigma})} 
\label{eq214}
\end{equation}
\\
with the $\pi\pi$ and $\sigma\sigma$ bubbles given by       
\\
\ba
i\Sigma_{\pi\pi}(p^2) &=& 
\int \frac{d^{4}q}{(2\pi)^{4}} \frac{1}{q^2-{\cal E}_{\pi}^2 +i\eta }\,\, 
\frac{1}{(p-q)^2-{\cal E}_{\pi}^2 +i\eta }
\nonumber\\
i\Sigma_{\sigma\sigma}(p^2)& = &\int \frac{d^{4}q}{(2\pi)^{4}}
\frac{1}{q^2-{\cal E}_{\sigma}^2 +i\eta }\,\,
 \frac{1}{(p-q)^2-{\cal E}_{\sigma}^2 +i\eta} 
\label{eq215}
\ea
\\
It should be mentioned that the condensate $s$ does not receive any 
contributions from RPA fluctuations in this approximation. 

In spite of the fact that the Goldstone theorem is fulfilled by the 
formalism described above it has a severe draw back
for the $\pi-\pi$ scattering problem. As mentioned in the introduction
there will be interactions in the chiral limit due to the fact that 
the quasipions remain massive.\\ 
It is now interesting to consider the limit ($N \rightarrow \infty)$. By 
inspection of eqs.~(\ref {eq207}),(\ref {eq212})and (\ref {eq214}) the 
HFB-RPA approach yields the following results
\\ 
\begin{eqnarray}
 \frac{c}{s} &\,=\,&  \mu_0^{2} + 
   \lambda_0^{2} \left[ I_{\pi}  + s^{2} \right]
 \nonumber\\
 m_{\pi}^2 &\,=\,& {\cal E}_{\pi}^2 = \frac{c}{s}
 \nonumber\\
 {\cal E}_{\sigma}^2&=& \frac{c}{s} + 2 \lambda_0^2 s^2 
\nonumber\\
m^2_{\sigma} &\,=\,& {\cal E}^2_{\sigma} \quad+\quad
2 \lambda_0^4 s^2
\frac{ \Sigma_{\pi\pi}(m^2_{\sigma})}
{\left[ 1\,-\,  \lambda_0^2 \Sigma_{\pi\pi}(m^2_{\sigma})\right]}
 \label{eq216}
 \end{eqnarray}
\\
where the $\Sigma_{\pi\pi}$ bubble is  built out of the asymptotic 
states
\\ 
\begin{equation}
\Sigma_{\pi\pi}(p^2) = 
-i\int \frac{d^{4}q}{(2\pi)^{4}} D_{\pi}(q) D_{\pi}(p-q)
 \label{eq217}
 \end{equation}
\\
which are 'acceptable' from the point of view of chiral symmetry. 
Obviously the pion is now a Goldstone boson already at the mean-field 
level. The RPA fluctuations do not survive in the large-$N$ limit in the 
case of the pion while for the sigma they are still partly present. In fact, 
as will become clear in sect.~3, the result obtained for the pion mass 
corresponds exactly to the Hartree-Bogoliubov solution since terms
induced by the Bose statistics disappear. The above results have been 
obtained previously by several authors using various methods \cite{BG,Abot}.
What we have presented so far is yet another way of deriving these results.
They merely serve as a motivation for introducing a systematic $1/N$ 
expansion in the next section. 

First we wish to address another issue, however. In contrast to the HBF-RPA 
approach in $O(4)$ the large-$N$ limit permits a relatively 
straightforward regularization. The bare coupling constant $\lambda_0$ 
and bare mass $\mu_0$ are replaced by renormalized quantities: 
\\ 
\begin{equation}
\lambda^2 \,=\, \frac{\lambda_0^2}{1-\lambda_0^2 L_0(\Lambda)}
\quad\quad\quad 
\mu^2 \,=\,\mu_0^2(1+\lambda^2 L_0(\Lambda)) \,+\, \lambda^2K_0(\Lambda) 
\label{eq218}
\end{equation}
\\ 
where $L_0$ and $K_0$ represent a logarithmic divergence 
and a quadratic divergence, respectively,
and $\Lambda$ is a four-dimensional cut-off. Defining the regularized pionic 
tadpole and the $\pi\pi$ bubble as
\\ 
\begin{equation}
\bar{I}_{\pi} \, = \, I_{\pi} - K_0(\Lambda) - m_{\pi}^2 L_0(\Lambda) 
\quad\quad\quad 
\bar{\Sigma}_{\pi\pi}(p^2)\, = \, {\Sigma}_{\pi\pi}(p^2) - L_0(\Lambda) 
\label{eq219}
\end{equation}
\\ 
the physical pion- and sigma masses are rendered finite
\\ 
\begin{eqnarray}
 m_{\pi}^2 &\,=\,& \frac{c}{s} \,=\,  \mu^{2} + 
   \lambda^{2} \left[ \bar{I}_{\pi}  + s^{2} \right]
 \nonumber\\
m^2_{\sigma} &\,=\,& \frac{c}{s} \,+\,   
\frac{2 \lambda^2 s^2}
{\left[ 1\,-\,  \lambda^2 \bar{\Sigma}_{\pi\pi}(m^2_{\sigma})\right]}
 \label{eq220}.
 \end{eqnarray}
\\
In addition, the well-known Ward identity for the $\sigma\pi\pi$ vertex
\cite{BLEE} is obeyed:  
\\ 
\begin{equation}
m^2_{\sigma} \,-\, m_{\pi}^2 \,=\,    
\frac{2 \lambda^2 s}
{\left[ 1\,-\,  \lambda^2 \bar{\Sigma}_{\pi\pi}(m^2_{\sigma})\right]}\, s  .
 \label{eq221}
 \end{equation}
\\
A comment is in order here. From eq.~(\ref{eq218}) one can 
easily see that the renormalized coupling constant $\lambda$ vanishes  
when the cut-off $\Lambda$ is removed and hence the theory becomes trivial. 
The triviality of the $\lambda \Phi^4$-theory in the $1/N$ approach has been 
thoroughly studied (see for instance \cite{Abot,Root}) and, to leading 
order, it has been shown in \cite{Abot} that the vacuum has the same group of
invariance as the Lagrangian. The same is evidently true here. This problem may 
be avoided, however,  by assigning a physical significance to a finite 
cut-off $\Lambda$, 
arguing that the sigma-model only describes physics up to a certain scale.
In this spirit it was shown in ref.~\cite{Nune} that, with a cutoff version 
of the  $\lambda \Phi^4$-theory, the nontrivial vacuum can be stabilized.
Another interesting possibility is the introduction of more degrees of
freedom (vector bosons in the case of $\lambda \Phi^4$ \cite{OlSCHN})
which may have its own difficulties, however. Clearly more work will be 
needed on this point.

\section{Dynamical Boson Mapping and the $1/N$-Expansion}
 
In this section we present an approach to the $1/N$-expansion in the 
cut-off version of the $O(N+1)$ sigma-model inspired by well-known boson 
mapping techniques used in nuclear physics \cite{BLRI, RS80, M93}.
This method is systematic in the sense that it represents a well-defined 
expansion in powers of $1/ \sqrt{N}$ and no a posteriori limit 
$N\rightarrow \infty$ has to be taken.
In fact, the RPA is based on the so-called quasi-boson approximation which,
in the nuclear physics context, means that Fermion pairs are replaced by ideal
bosons. Analogously the RPA for bosons, as applied in the last section, 
implies the  bosonization of boson pairs. Boson expansion techniques for 
interacting bose systems have first been studied by Curutchet, Dukelsky, Dussel, and Fendrik (CDDF) in \cite{CDDF} and by 
  Bijker, Pittel and 
Dukelsky (BPD) in \cite{PIDUK}. We will apply this technique to obtain a 
systematic $1/N$ expansion of the Hamiltonian of the linear sigma-model.\\
It has been discussed at length by Marshalek et al. \cite{M74, M93} how such 
an expansion preserves at the same time the symmetries order by order. 
We are specifically interested in the $\pi-\pi$ scattering problem for which 
we want to establish a Lippmann-Schwinger type of equation which fulfills 
unitarity as well as constraints from chiral symmetry such as the vanishing 
of the $\pi-\pi$ scattering length once the explicit symmetry breaking
is removed $(m_{\pi}\rightarrow 0)$.
To proceed one introduces ideal bose operators $ (A^+_{q,p},  A_{q,p})$ 
and bosonizes pion pair operators as first laid out by Holstein and Primakoff
\cite{HOLPRI} (in nuclear physics it has become known also as the 
Belyaev-Zelevinsky method \cite{BEZE}). The method consists of the 
following mapping 
\\
\begin{eqnarray}
{\vec{a}}^{+}_{q}{\vec{a}}^{+}_{p} &\, = \,& \left( A^+\sqrt{ N \,+\, A^+A } \right)_{q,p}\nonumber\\ 
{\vec{a}}_{q}{\vec{a}}_{p} &\, = \,& \left( {\vec{a}}^{+}_{p}{\vec{a}}^{+}_{q} \right)^+ \nonumber\\
{\vec{a}}^{+}_{q}{\vec{a}}_{p}  &\, = \,& \left( A^+ A \right)_{q,p} 
\label{eq301}
\end{eqnarray}
\\
where the new operators $(A^+_{q,p}, A_{q,p})$ are ideal bosons and 
obey the usual Heisenberg-Weyl algebra:
\\
\begin{eqnarray}
\left[ A_{m,n} \, , \, A^+_{p,q} \right] &\,=\,&  \left( \delta_{m,q} \delta_{n,p} + \delta_{n,q} \delta_{m,p} \right) \nonumber\\
\left[A^{+}_{m,n} \, , \, A^{+}_{p,q} \right] & \,=\, & \left[ A_{m,n} 
\, , \, A_{p,q}\right] \, =\, 0.
\label{eq302}
\end{eqnarray}
\\
With respect to the more familiar Fermion case one should notice the change 
of sign on the {\sl rhs} of the first equation in eq.~(\ref{eq302}) and under 
the square root in eq.~(\ref{eq301}). In addition, the operators $A_{q,p}$ 
are symmetric under exchange of indices  
\\
\begin{equation}
 A_{q,p} \,=\, A_{p,q}\quad\quad\quad   A^+_{q,p} \,=\, A^+_{p,q}
\label{eq303}
\end{equation}
\\
rather than antisymmetric as in the case of Fermion bosonization.
The approach of Belyaev-Zelevinski to the bosonic mapping requires the 
realization of the original algebra for pairs of pions 
\\
\begin{eqnarray}
\left[{\vec{a}}_{1}{\vec{a}}_{2}\,\, ,\,\, {\vec{a}}^{+}_{3}{\vec{a}}^{+}_{4} \right] &\,=\,& 
 N \left( \delta_{13} \delta_{24} +  \delta_{14} \delta_{23} \right) \, +\,
 \delta_{13}{\vec{a}}^{+}_{4}{\vec{a}}_{2} +  
 \delta_{23}{\vec{a}}^{+}_{4}{\vec{a}}_{1} +
 \delta_{14}{\vec{a}}^{+}_{3}{\vec{a}}_{2} +
 \delta_{24}{\vec{a}}^{+}_{3}{\vec{a}}_{1} \nonumber\\
\left[ {\vec{a}}_{1}{\vec{a}}_{2}\,\,\, ,\,\,\, {\vec{a}}^{+}_{3}{\vec{a}}_{4} \right] &\,=\,&
 \delta_{13}{\vec{a}}_{2}{\vec{a}}_{4} +  
 \delta_{23}{\vec{a}}_{1}{\vec{a}}_{4} \nonumber\\
\left[{\vec{a}}^{+}_{1}{\vec{a}}^{+}_{2}\, ,\, {\vec{a}}^{+}_{3}{\vec{a}}^{+}_{4} \right] &\,=\,&
 \left[{\vec{a}}_{1}{\vec{a}}_{2}\, ,\, {\vec{a}}_{3}{\vec{a}}_{4} \right]
 \, =\,0.
\label{eq304}
\end{eqnarray}
\\
The reader may verify that this is in fact true order by order, using 
the Holstein-Primakoff mapping in  eq.~(\ref{eq301}).\\
To write the Hamiltonian in the (CDDF-BPD)-representation we need to define the
sigma field. Since the latter is to be constructed perturbatively, it is
legitimate to use a form similar to eq.~(\ref{eq202})
\\
\begin{equation}
  {\hat \sigma}({\bf x}) \,=\,  \int \frac{d^3q}
  {\sqrt{(2\pi)^{3} 2{\cal E}_{\sigma}(q)}}
  \left(b_{{\bf q}} e^{i{\bf q}{\bf x}} + b^{+}_{{\bf q}}
  e^{-i{\bf q}{\bf x}} \right)
\label{eq305}
\end{equation}   
\\
with the sigma-quasiparticle mass ${\cal E}_{\sigma}$ to be defined later.
With these definitions the Hamiltonian takes the following form:
\\
 \begin{eqnarray}
    H &=& {\cal H}_{0} \,+\, 
 \int\frac{d^3q}{2{\cal E}_{\sigma}(q)} \left[ \left({\cal E}^2_{\sigma}(q) +
\omega_q^2 +\lambda_0^2 I_0  \right) b_q^+b_q - 
\frac{1}{2} \left( {\cal E}^2_{\sigma}(q) - \omega_q^2 -\lambda_0^2 I_0  \right)\left(  b_q^+b^+_{-q} +  b_{-q}b_{q}\right) \right]     \nonumber\\
  \,\,&+&\,\,
  \int  d^3q \quad  e_0(q) \left(A^+A\right)_{q,q} \,+\,
  \frac{\Delta_0(q)}{2} \left[ \left(A^+\sqrt{ N\,+\,A^+A}\right)_{q,-q}
 \,+\, \left(\sqrt{ N\,+\,A^+A}A\right)_{-q,q}  \right] \nonumber\\
 &+& 
  \int d{\bf x} \,\,\, \left[
\frac{\lambda_0^2}{4 N}\left( {\bf{:\,{\vec \pi}^2(x)\,:}} +
{\hat \sigma}^2(x) \right)^2
 \,-\, \sqrt{N} c {\hat \sigma}(x) \right]
\label{eq306}
\end{eqnarray}
\\
where the colon indicates normal ordering and the various functions 
are given by
\\
\begin{eqnarray}
{\cal H}_{0} \,&=&\, \frac{N \lambda_0^2}{4} I_0^2 
\quad\quad with \quad\quad I_0\,=\,\int\frac{d^3q}{(2\pi)^{3}}\frac{1}
{2\omega_q} 
\nonumber\\
e_0(q) \,&=&\, \omega_q \,+\, \Delta_0(q),  \quad\quad\quad\quad\quad
 \Delta_0(q)\,=\, \frac{ \lambda_0^2}{2\omega_q} I_0.
\label{eq307}
\end{eqnarray}
\\ 
The normal-ordered pion fields can be explicitly written as
\\
\begin{equation}
:\,{\vec \pi}^2({\bf x}): \,=\, \int  \frac{d^3q_1 d^3q_2
 e^{i({\bf q}_1 +{\bf q}_2) {\bf x}}
 }{(2\pi)^{3} \sqrt{4 \omega_1\omega_2}} 
 \left[ \left(A^+\sqrt{ N\,+\,A^+A}\right)_{1,2}
 \,+\, \left(\sqrt{ N\,+\,A^+A}A\right)_{-1,-2} \,+\, 2 
 \left(A^+A\right)_{1,-2}  \right] .
\label{eq308}
\end{equation}
\\      
Clearly the Hamiltonian is an infinite series in powers of the new bosons
as a direct result of the Holstein-Primakoff mapping. It then takes the 
form
\\
\begin{equation}
H \,\,=\,\, {\cal H}^{(0)}\,+\,{\cal H}^{(1)}\,+\,{\cal H}^{(2)}\,+\,{\cal H}^{(3)}\,+\,{\cal H}^{(4)}\,+...
\label{eq309}
\end{equation}
\\      
The superscript in each term designates the power in the boson operators 
and these are either the sigma bosons $b_q, b_q^+$ or the newly introduced 
pion-pair bosons $A, A^+$. In fact, in addition to this expansion in the 
power of bosons there exists an underlying expansion in the square root of 
the number of pion charges, $\sqrt{N}$. The latter is particularly suitable 
for a perturbative treatment of the dynamics generated by the Hamiltonian 
eq.~(\ref {eq306}). The fact that $N$ is arbitrary insures the preservation
of the symmetries of $H$ at each order. The perturbative calculation of all
higher orders is performed in a selfconsistent basis which we will 
establish first. Since there are two bosons which are able 
to condense in the vacuum, the natural choice for the trial wave function 
of the ground state is a coherent state of the form
\\
\begin{equation}
\ket{ \psi} = exp\left[ \sum_{q} {\hat d}_q \,A^+_{q,-q} \,\, + \,\, 
 \ave{b_0}\, b_{0}^+ \right] \ket{ 0}.
\label{eq310}
\end{equation}
\\
To fix the parameters $ {\hat d}_q$ and $\ave{b_0}$ one needs to 
apply the Ritz variational principle to the expectation value 
$ {\cal H}^0 =\frac{\bra{\psi } H \ket{ \psi }}
{\ave{\psi\mid \psi} }$. The latter can be evaluated and gives
\\
\begin{equation}
 {\cal H}^{0} \,=\, (2\pi)^{3} \, N \,\left[ 
\int \frac{d^3 {\bf q}}{(2\pi)^{3}} \omega_q \frac{{\hat d}_q^2}{N} \quad+\, 
\frac{\lambda_0^2 \ave{\hat \sigma}^2}{2N}\left(I_0 + K_0\right) \,+\,
\frac{\lambda_0^2}{4}\left(I_0 + K_0\right)^2 \,+\, 
\frac{\lambda_0^2 \ave{\hat \sigma}^4}{4 N^2} \,+\,
\frac{\mu_0^2 \ave{\hat \sigma}^2}{2N}\,-\, \frac{c\ave{\hat \sigma}}
{\sqrt{N}}  \right]
\label{eq311}
\end{equation}  
The parameter  $\ave{\hat \sigma}= \frac{\ave{b_0}+\ave{b_0^+}}{\sqrt{(2\pi)^3 2{\cal E}_{\sigma}}} $, as defined in eq.~(\ref{resq1}),  
is nothing but the expectation value of the sigma field 
in the coherent state $\ket{ \psi }$ while $K_0$ is given by
\\
\begin{equation}
K_0\,=\, \frac{1}{N} \, \int  \frac{d^3 {\bf q}}{(2\pi)^{3}} 
\frac{\left({\hat d}_q\,+\,\sqrt{N\,+\, {\hat d}_q^2}\right)^2\,-\, N}
{2\omega_q}.
\label{eq312}
\end{equation}
\\
The minimum of $H^0$ determines the values of  ${\hat d}_q$ and 
$s\,=\, \ave{\hat \sigma}/\sqrt{N}$, as solutions of the following 
equations
\\
\begin{eqnarray}
  e_{\pi}(q)\left(2\, {\hat d}_q \,\sqrt{N\,+\, {\hat d}_q^2}\right) 
  \,&+&\, \Delta_{\pi}(q)\left(2\,{\hat d}_q^2 \,+\,N \right) \,=\, 0
\nonumber\\
\mu_0^2\,&+&\, 2\,\omega_q\,\Delta_{\pi}(q) \,=\, \frac{c}{s}
\label{eq313}
\end{eqnarray}
\\ 
where $\Delta_{\pi}(q)$ and $e_{\pi}(q)$ are given by
\\
\begin{equation}
e_{\pi}(q)\,=\,\omega_q\,+\, \Delta_{\pi}(q), \quad\quad\quad
\Delta_{\pi}(q)\,=\, \frac{\lambda_0^2}{2\,\omega_q}
\left[ I_0\,+\, K_0\,+\, s^2\right].
\label{eq314}
\end{equation}
\\
The first equation in (\ref{eq313}) can be reexpressed in the more 
suggestive form
\\
\begin{equation}
{\hat d}_q^2 \,=\, \frac{N}{2}\left[\frac{e_{\pi}(q)}
{\sqrt{e_{\pi}^2(q)\,-\,\Delta_{\pi}^2(q)}}\,-\, 1 \right].
\label{eq315}
\end{equation}
\\
From the canonical form of this 'gap equation' one can deduce 
the pion quasiparticle energy ${\cal E}_{\pi}(q)$ as
\\
\begin{equation}
{\cal E}_{\pi}(q) = \sqrt{e_{\pi}^2(q)\,-\,\Delta_{\pi}^2(q)} \,=\,
\sqrt{q^2 \,+\, \mu_0^2 \,+\,\lambda_0^2\left[I_0\,+\,K_0\,+\,s^2\right] }.
\label{eq316}
\end{equation}
\\
One can also deduce from eq.~(\ref{eq314}) that the shift ${\hat d}_q$ for 
the pion pairs can have a simple scaling, namely as $\sqrt{N}$, which shows  the physical 
similarity of this parameter to the scalar field condensate 
$\ave{\hat \sigma}$. In what follows we will use a rescaled shift parameter
\[
  d_q \,=\, {\hat d}_q/{\sqrt{N}}.
\]
An interesting identity which allows to transmit the selfconsistency of the 
gap equation in eq.~(\ref {eq314}) to the pion quasiparticle mass reads
\\
\begin{equation}
{\cal E}_{\pi}(q) = \left( d_q \,-\,\sqrt{1\,+\,d_q^2} \right)^2 \, 
\omega_q 
\label{eq317}
\end{equation}
\\
from which one can finally extract the Hartree-Bogoliubov (HB) mass of the 
pion,
\[
  m_{\pi} = {\cal E}_{\pi}(0), 
\]
and the value of the condensate via the 
two coupled equations
\\
\begin{eqnarray}
 m_{\pi}^2&\, =\,& \mu_0^{2} +  \lambda_0^{2} \left[ I_{\pi}  + 
 s^{2} \right]
 \nonumber\\
\frac{c}{s} &\,=\,&  \mu_0^{2} + 
   \lambda_0^{2} \left[ I_{\pi}  + s^{2} \right]
\label{eq318}
\end{eqnarray}
\\
where $I_{\pi}= I_0 + K_0 $ denotes the quasi-pion tadpole.\\ Having fixed the 
quasiparticle  basis we can now expand the Hamiltonian in eq.~(\ref{eq306})
in powers of the shifted bosons operators $ \tilde {A}_{q,p}$ and 
$\beta_q$ defined as
\\    
\begin{eqnarray}
\tilde {A}_{q,p}\,=\, A_{q,p}\,&-&\, \sqrt{N}d_q \delta(q+p) \quad\quad\quad
\beta_q\,=\, b_q \,-\, \ave{b_0}\, \delta(q) \nonumber\\
with\quad\quad \quad\quad \tilde{A}\ket{ \psi}&=& \beta \ket{ \psi}\,=\, 0.  
\label{eq319}
\end{eqnarray}
\\
Knowing the scaling of all parameters in the Hamiltonian the latter can 
therefore be expanded without ambiguity according to
\begin{equation}
H \,\,=\,\, N\,H^{(0)}\,+\,\sqrt{N}\,H^{(1)}\,+\,H^{(2)}\,+
\,\frac{1}{\sqrt{N}}\,H^{(3)}\,
+\,\frac{1}{N}\,H^{(4)}\,+...
\label{eq320}
\end{equation}
\\
where each order is obtained by expanding the square roots in 
eq.~(\ref{eq301}). Using the parameter differentiation techniques of 
operators (see appendix 1) one can write down the contributions to the
 Hamiltonian to leading orders in  $1/N$, namely $H^{(1)}$ and $H^{(2)}$:
\\
\begin{eqnarray} 
H^{(1)} &=& 
\sqrt{ \frac{(2\pi)^3}{2{\cal E}_{\sigma}}} \,\,\left[ \lambda_0^{2} I_{\pi}s  \,+\, \lambda_0^{2} s^{3}\,+\,\mu_0^2s \,-\,c \right]\left( \beta_0 \,+\, \beta^+_0 \right)
\nonumber\\
\quad\quad\quad\quad &+&\quad\quad\quad
 \int d^3q \,\left[ \,\omega_q d_q \,+\,
\frac{\Delta_{\pi}(q)\, \left(d_q \,+\, \sqrt{1\,+\,d_q^2}\right)^2}{ 2\sqrt{1\,+\,d_q^2}}\,
\right]\,\,\left(\tilde{A}^+_{q,-q}\,+\,\tilde{A}_{-q,q}\right)
\label{eq321}
\end{eqnarray}
\\
\begin{eqnarray} 
H^{(2)} &=& \int d^3q {\cal E}_{\sigma}(q) \beta^+_q\beta_q \quad+\quad
 \int d^3q {\cal E}_{\pi}(q)\left(1\,+\, d_q^2 \right)
\left(\tilde{A}^+\tilde{A}\right)_{q,q}
 \nonumber\\
\nonumber\\
&+& \int d^3q d^3p \left[ 
\frac{\Delta_{\pi}(q)}{2}\frac{\sqrt{1+d_q^2}-\sqrt{1+d_p^2}}{d_q^2-d^2_p}
\left(\left(d\tilde{A} + \tilde{A}^+d\right)_{q,p}\tilde{A}_{p,q}\,+\,
\tilde{A}^+_{q,p} \left(d\tilde{A} + \tilde{A}^+d\right)_{p,q} \right)
\right.
\nonumber\\
\nonumber\\
&+& 
\left.
\frac{d_q\,\Delta_{\pi}(q)}{d_q^2\,-\,d_p^2}\left(\frac{1}{2\sqrt{1+d_q^2}}
 - \frac{\sqrt{1+d_q^2}-\sqrt{1+d_p^2}}{d_q^2-d^2_p}\right)
\left(d\tilde{A} + \tilde{A}^+d\right)_{q,p}
\left(d\tilde{A} + \tilde{A}^+d\right)_{p,q} \right]\nonumber\\
\nonumber\\
&+& 
\lambda_0^2\,s
\, \int\frac{d^3q_1d^3q_2d^3q_3 }{\sqrt{(2\pi)^3\, 8 {\cal E}_{\sigma}
(q_3)\omega(q_1)\omega(q_2)}}\,\,  \delta^3(q_1+q_2+q_3) \Gamma_{1,2}
\left[\beta_3\,+\,\beta^+_{-3}\right]\left[ \tilde{A}_{1,2}\,+\,
\tilde{A}^+_{-1,-2}\right]\nonumber\\
\nonumber\\
&+&\frac{\lambda_0^2}{4} 
\int\frac{d^3q_1d^3q_2d^3q_3 d^3q_4 \quad \delta^3(q_1+q_2+q_3+q_4)   }{(2\pi)^3\sqrt{16 \omega(q_1)
\omega(q_2)\omega(q_3)\omega(q_4)}}\,
\,\Gamma_{1,2}\,\Gamma_{3,4}\,
\left[ \tilde{A}_{1,2}+\tilde{A}^+_{-1,-2}\right]
\left[ \tilde{A}_{3,4}+\tilde{A}^+_{-3,-4}\right]\nonumber\\
\label{eq322}
\end{eqnarray}
\\
with the following definitions for the sigma quasiparticle mass 
${\cal E}_{\sigma}$ and for the factors $\Gamma_{i,j}$:
\\
\begin{eqnarray}
{\cal E}_{\sigma}^2 \,&=&\, m_{\pi}^2 \,+\, 2\lambda_0^2 s^2
\nonumber\\
\Gamma_{i,j} \,&=&\, \frac{\left(d_i+\sqrt{1+d_i^2}\right)^2-\left(d_j
+\sqrt{1+d_j^2}\right)^2}{2\left(d_i\,\,-\,\,d_j\right)}.
\label{eq323}
\end{eqnarray}
\\
The gap equation which we have derived in eq.~(\ref{eq313}) eliminates 
the linear part of the Hamiltonian, {\sl i.e} $ H^{(1)}\,=\,0$ thus defining 
the HB basis. The leading contribution in the expansion is therefore of
order one. As we have seen, 
 an appropriate mapping of the boson pair operators into ideal 
bose operators automatically  produces the correct result. \\
The Hamiltonian above can be recast in a more compact form which is 
particularly suitable for the forthcoming considerations. To achieve this 
a set of new operators $B^+_{q,p},B_{q,p}$ is defined which corresponds to
a rotation of the set of $\tilde{A}^+_{q,p},\tilde{A}_{q,p}$ operators, 
according to
\\
\begin{equation}
B^+_{q,p}\,=\, G_{q,p}\,\tilde{A}^+_{q,p}\,\,-\,\,H_{q,p}\,\tilde{A}_{q,p}
\label{eq324}
\end{equation}
\\
In order for the $B^+_{q,p},B_{q,p}$ operators to obey the same algebra as 
the $\tilde{A}^+_{q,p},\tilde{A}_{q,p}$, the coefficients $G_{q,p}$ and 
$ H_{q,p}$ have to be chosen such that
\\
\begin{equation}
   G_{q,p}^2 \,\,\,-\,\,\, H_{q,p}^2 \,=\,1 .
\label{eq325}
\end{equation}
\\
Explicitly the functions $G_{q,p}$ and $ H_{q,p}$ are given by
\\
\begin{equation}
   G_{q,p}\,\pm\, H_{q,p} \,=\, \frac{d_q\sqrt{1\,+\,d_p^2}\,
   \pm\,d_p\sqrt{1\,+\,d_q^2}}{d_q\,\,\pm\,\,d_p} .
\label{eq326}
\end{equation}
\\
With these definitions one verifies that the symmetry of the operators 
$\tilde{A}^+_{q,p},\tilde{A}_{q,p}$ with respect to the interchange of the 
indices is also obeyed by the new bosons $B^+_{q,p},B_{q,p}$, since  
$ G_{q,p}$ and $ H_{q,p}$ are symmetric under the interchange of the 
indices. The quadratic part of H now reads
\begin{eqnarray}
H^{(2)} &=& \int d^3q\,\, {\cal E}_{\sigma}(q)\,\, \beta^+_q\beta_q \quad+\quad
\int d^3q d^3p\,\, {\cal E}_{\pi}(q)\,\,B^+_{q,p}B_{q,p}
\nonumber\\
&+& 
\lambda_0^2\,s
\, \int\frac{d^3q_1d^3q_2d^3q_3 \quad \delta^3(q_1+q_2+q_3)  }{\sqrt{(2\pi)^3\, 8 {\cal E}_{\sigma}
(q_3){\cal E}_{\pi}(q_1){\cal E}_{\pi}(q_2)}}\,\,  
\left[\beta_3\,+\,\beta^+_{-3}\right]\left[ B_{1,2}\,+\,B^+_{-1,-2}\right]
\nonumber\\
&+&\frac{\lambda_0^2}{4} 
\int\frac{d^3q_1d^3q_2d^3q_3 d^3q_4 \quad \delta^3(q_1+q_2+q_3+q_4)}{(2\pi)^3\sqrt{16 
{\cal E}_{\pi}(q_1){\cal E}_{\pi}(q_2)
{\cal E}_{\pi}(q_3){\cal E}_{\pi}(q_4)}}\,
\,\,
\left[ B_{1,2}\,+\,B^+_{-1,-2}\right]
\left[ B_{3,4}\,+\,B^+_{-3,-4}\right].\nonumber\\
\label{eq327}
\end{eqnarray}
\\
The next task is to diagonalize each order of the Hamiltonian. Parts with odd 
powers in the 
operators have to be eliminated by this procedure, as was the case for 
$H^{(1)}$. Expressing $H^{(2)}$ in diagonal form leads to an 
expression with uncoupled oscillators as will be seen below.\\
The diagonalization of $H^{(2)}$ can be performed by using a general 
Bogoliubov rotation which mixes the configuration of the quasisigma Boson 
and the bosonized pair of pions according to
\\
\begin{equation}
Q^+_{\nu}({\vec p})\,=\, U^{(1)}_{\nu}({\vec p})\beta^+_{\vec p}\,
-\, V^{(1)}_{\nu}({\vec p})\beta_{-{\vec p}}
\,+\,\frac{1}{\sqrt{2}} \sum_{q} \left[ U^{(2)}_{\nu}({\vec p},{\vec q})B^+_{{\vec q},{\vec p}
-{\vec q}}\,-\, V^{(2)}_{\nu}({\vec p},{\vec q})B_{-{\vec q},-{\vec p}+{\vec q}}\right].
\label{eq328}
\end{equation} 
\\
The operator $Q_{\nu}^+$ can be considered as an RPA excitation operator. 
The rotation in the RPA operator in eq.~(\ref{eq328}) is performed under the 
kinematical constraint $\delta({\vec q}_1 + {\vec q}_2 -{\vec p})$,
where $\vec p$ denotes the total momentum.\\ 
Using Rowe's equation of motion \cite{ROW} 
\\
\begin{equation}
 \bra{RPA } \left[ \delta Q_{\nu}({\vec p}) \, \, , \, \, \left[H^{(2)}\, , \, Q_{\nu}^+({\vec p}) \right] \right] \ket{ RPA} = \Omega_{\nu}({\vec p})
 \bra{RPA } \left[ \delta Q_{\nu}({\vec p})  \, , \, Q_{\nu}^+({\vec p})
 \right] \ket{ RPA}
\label{eq329}
\end{equation} 
\\
and keeping in eq.~(\ref{eq329}) the full RPA ground state defined by 
$Q_{\nu}\ket{RPA}=0$ would lead to a selfconsistent form of the RPA 
equations (see \cite{DUSCH}) if higher than second-order terms are included 
in the expansion of $H$. Retaining only $H^{(2)}$ linearizes the 
equation of motion  and one obtains the familiar form of the RPA 
equations
\\                                            
\begin{equation}
\int d^3 {\vec q}_2
\left(
\begin{array}{cc}
{\cal A}_{\vec p}({\vec q}_1,{\vec q}_2) &
{\cal B}_{\vec p}({\vec q}_1,{\vec q}_2) \\ 
{\cal B}_{\vec p}({\vec q}_1,{\vec q}_2) &
{\cal A}_{\vec p}({\vec q}_1,{\vec q}_2)
\end{array} 
\right) 
\left(\begin{array}{c}
{\cal U}_{\nu}({\vec p},{\vec q}_2) \\ {\cal V}_{\nu}({\vec p},{\vec q}_2)
\end{array} \right) 
\, =\, \Omega_{\nu}({\vec p}) {\cal N}
\left(\begin{array}{c}
{\cal U}_{\nu}({\vec p},{\vec q}_1) \\ {\cal V}_{\nu}({\vec p},{\vec q}_1)
\end{array}\right)
\label{eq330}
\end{equation}
\\
where ${\cal A}_{\vec p}$, ${\cal B}_{\vec p}$ are $2\times 2$ matrices 
given by
\\
\begin{equation}
{\cal A}_{\vec p}({\vec q}_1,{\vec q}_2) = 
\left(
\begin{array}{cc}
{\cal E}_{\sigma}({\vec p}) & 0 \\
0 & \left[{\cal E}_{\pi}({\vec q}_1) + {\cal E}_{\pi}({\vec p}-{\vec q}_1)
\right]
\end{array} 
\right) \, \delta({\vec q}_1-{\vec q}_2)
\quad +\quad  {\cal B}_{\vec p}({\vec q}_1,{\vec q}_2) 
\end{equation}
with 
\begin{equation}
{\cal B}_{\vec p}({\vec q}_1,{\vec q}_2) = 
\left(
\begin{array}{cc}
 0  &
\frac{\sqrt{2}  \lambda^2_0 s}{\sqrt{(2\pi)^3 8 {\cal E}_{\sigma}({\vec p}) {\cal E}_{\pi}({\vec q}_2) {\cal E}_{\pi}({\vec p}-{\vec q}_2)}} \\
\frac{ \sqrt{2} \lambda^2_0 s}{\sqrt{(2\pi)^3 8 {\cal E}_{\sigma}({\vec p}) {\cal E}_{\pi}({\vec q}_1) {\cal E}_{\pi}({\vec p}-{\vec q}_1)}} &
\frac{ \lambda^2_0 }{(2\pi)^3 \sqrt{ 16 {\cal E}_{\pi}({\vec q}_1) 
{\cal E}_{\pi}({\vec q}_2) {\cal E}_{\pi}({\vec p}-{\vec q}_1)        
{\cal E}_{\pi}({\vec p}-{\vec q}_2)}}
\end{array} 
\right) 
\label{eq331}
\end{equation}
\\
respectively. The $4\times 4$ norm matrix ${\cal N}$ as well as the 
two-component vectors  ${\cal U}_{\nu}$, ${\cal V}_{\nu}$ are given by
\\
\begin{eqnarray}
{\cal N} &=& \left(\begin{array}{cc}
I_d & 0 \\ 0 & -I_d
\end{array} \right) \quad\quad 
I_d  = \left(\begin{array}{cc}
1 & 0 \\ 0 & 1 \end{array} \right)\nonumber\\
\nonumber\\ 
{\cal U}_{\nu}({\vec p},{\vec q}) &=& 
\left(
\begin{array}{c}
 U^{(1)}_{\nu}({\vec p})  \\  U^{(2)}_{\nu}({\vec p},{\vec q})
\end{array} 
\right) \quad\quad
{\cal V}_{\nu}({\vec p},{\vec q}) = 
\left(
\begin{array}{c}
 V^{(1)}_{\nu}({\vec p})  \\  V^{(2)}_{\nu}({\vec p},{\vec q})
\end{array} 
\right).
\label{eq332}
\end{eqnarray}
\\
One can now solve the eigenvalue problem and extract the RPA frequencies
to obtain
\\
\begin{equation}
\Omega^2_{\nu}({\vec p}) \,=\, {\cal E}_{\sigma}^2 \,+\,   
\frac{2 \lambda_{0}^4 s^2\,{\Sigma}_{\pi\pi}(\Omega^2_{\nu}({\vec p}))}
{ 1\,-\,  \lambda_{0}^2 {\Sigma}_{\pi\pi}(\Omega^2_{\nu}({\vec p}))}\,+\,{\vec p}^2
\label{eq333}
\end{equation} 
\\
with ${\Sigma}_{\pi\pi}(\Omega^2_{\nu}({\vec p}))$ being the Lorentz invariant
$\pi\pi$ self energy of the sigma given by
\\
\begin{equation}
{\Sigma}_{\pi\pi}(\Omega^2_{\nu}({\vec p}))\,=\, \int \frac{d^3q}{(2\pi)^3}\, 
\frac{{\cal E}_{\pi}(q)\,+\,{\cal E}_{\pi}(p-q) }{2\, {\cal E}_{\pi}(q)
{\cal E}_{\pi}(p-q)} \,\,\frac{1}{\Omega^2_{\nu}(p)\,-\,\left({\cal E}_{\pi}(q)
\,+\,{\cal E}_{\pi}(p-q)\right)^2}.
\label{eq334}
\end{equation}
\\
It should be noted that, in contrast to (\ref{eq215}), where the quasipion was 
massive in the chiral limit, here the  ${\Sigma}_{\pi\pi}$ self energy in (\ref{eq334}) is built on massless Goldstone pions. 
Using the usual orthonormalisation condition of the RPA states, 
{\sl i.e} $\bra{RPA}Q_{\nu}Q^+_{\nu'}\ket{RPA} = \delta_{\nu \nu'}$, one obtains the following 
solution for the RPA eigenvectors 
\\
\begin{eqnarray}
U^{(1)}_{\nu}({\vec p}) &=& \frac{(2\pi)^{\frac{3}{2}}}{\sqrt{2}}\frac{V_{\pi\pi\rightarrow\sigma}}{\Xi({\vec p},\Omega_{\nu})}\frac{1}{\sqrt{2{\cal E}_{\sigma}({\vec p})}}\frac{1}{\Omega_{\nu}-{\cal E}_{\sigma}({\vec p})}\nonumber\\
V^{(1)}_{\nu}({\vec p}) &=& -\frac{(2\pi)^{\frac{3}{2}}}{\sqrt{2}}\frac{V_{\pi\pi\rightarrow\sigma}}{\Xi({\vec p},\Omega_{\nu})}\frac{1}{\sqrt{2{\cal E}_{\sigma}({\vec p})}}\frac{1}{\Omega_{\nu}+{\cal E}_{\sigma}({\vec p})}\nonumber\\
U^{(2)}_{\nu}({\vec p},{\vec q}) &=& \frac{1}{2} \frac{V_{\pi\pi\rightarrow\pi\pi}}{\Xi({\vec p},\Omega_{\nu})}\frac{1}{\sqrt{4 {\cal E}_{\pi}({\vec q}){\cal E}_{\pi}({\vec p}-{\vec q})}}\frac{1}{\Omega_{\nu}-{\cal E}_{\pi}({\vec q})-{\cal E}_{\pi}({\vec p}-{\vec q})}\nonumber\\
V^{(2)}_{\nu}({\vec p},{\vec q}) &=& - \frac{1}{2} \frac{V_{\pi\pi\rightarrow\pi\pi}}{\Xi({\vec p},\Omega_{\nu})}\frac{1}{\sqrt{4 {\cal E}_{\pi}({\vec q}){\cal E}_{\pi}({\vec p}-{\vec q})}}\frac{1}{\Omega_{\nu}+{\cal E}_{\pi}({\vec q})+ {\cal E}_{\pi}({\vec p}-{\vec q})  }\nonumber\\
\label{eq335}
\end{eqnarray} 
\\ 
where $V_{\pi \pi \rightarrow \sigma}$ and 
$V_{\pi \pi \rightarrow \pi \pi}(\Omega_{\pi},{\vec p})$  stand 
for the $ \pi \pi \rightarrow \sigma$ and $ \pi \pi \rightarrow \pi \pi$ tree
level transition matrix respectively. These as well as the function $\Xi$ 
are given by 
\\
\begin{eqnarray}
\Xi(\Omega_{\nu}, {\vec p}) &=& \left(2\pi\right)^{3/2} \left[ \frac{ \Omega_{\nu}\, 
\left(V_{\pi \pi \rightarrow \sigma}\right)^2}{\left[\Omega_{\pi}^2-
{\cal E}^2_{\sigma}({\vec p})\right]^2}\,\,-\,\, \frac{1}{4}
\left(V_{\pi \pi \rightarrow \pi \pi}(\Omega_{\nu},{\vec p})\right)^2 
\frac{\partial \Sigma_{\pi\pi}(\Omega_{\nu}, {\vec p})}{\partial 
\Omega_{\nu}}\right]^{1/2}\nonumber\\
V_{\pi \pi \rightarrow \pi \pi}(\Omega_{\nu},{\vec p})&=& 2 \lambda_0^2
\frac{\Omega^2_{\nu}-{\cal E}_{\pi}^2({\vec p})}{\Omega^2_{\nu}-{\cal E}_{\sigma}^2({\vec p})},
\quad\quad\quad\quad
V_{\pi \pi \rightarrow \sigma }\,=\, 2\lambda_0^2 s .\nonumber\\
\label{eq336}
\end{eqnarray}
\\
This concludes the discussion of the leading order dynamics. 
The physical observables, at this order, are given by the expressions in eq.~(\ref{eq318}) for the pion mass and the condensate and by  eq. ~(\ref{eq333}) for  the sigma mass. These were the same results obtained at the end of the previous section.

In the next section some remaining subtleties in relation with 
the properties of the RPA approximation will be pointed out. 

\section{The $\pi-\pi$ Goldstone Mode and its Dynamics, Sum Rules.}

In the chiral limit the RPA solutions have to yield a spurious mode
 corresponding physically to the case of two non-interacting Goldstone
pions. This mode has to occur if the massless pions have zero total
and zero relative momentum. Such a solution is not manifest from 
expression (\ref{eq333}) which has been obtained by solving the RPA
equations in the sigma channel via Feshbach projection. In the chiral
limit and for vanishing total momentum $\vec p$ the RPA frequency 
will just become the sigma mass as to be expected and as can also be seen 
by comparing eq.~(\ref{eq333}) with eq.~(20). That a Goldstone
  mode ({\sl i.e.} a Goldstone mode for the two pions)  must also  exist,
 can be deduced from the following symmetry considerations
 (see also section 5).

Since, without explicit symmetry breaking, the sigma-model is invariant 
under chiral transformations, the Hamiltonian must fulfill the following 
commutation relations 
\\
\begin{equation}
\left[ H\,,\, Q_5^i \right] =  \left[ H\,,\, {\vec Q}_5{\vec Q}_5 \right] 
= 0
\label{eq411} 
\end{equation}
\\
since $ Q_5^i$, the i-th  axial charge, is a generator of the symmetry. The 
second relation is most interesting in the present context since it
involves two-pion degrees of freedom.
As for single $Q_5^i$ operators, the product of two axial charges
can also be mapped into the bosons $\tilde{A}^+,\tilde{A}$ or equivalently 
$B^+, B$. This allows to  express $ S \equiv {\vec Q}_5{\vec Q}_5 $ in an
infinite series in powers of the new bosons
\\
\begin{equation}
S \,\,=\,\,N^2\, S^{(0)}\,+\, N^{3/2}\,S^{(1)}\,+\,N\,S^{(2)}\,+ 
\,\sqrt{N}\,S^{(3)}\,+\,\,S^{(4)}\, 
+\,\frac{1}{\sqrt{N}}\,S^{(5)}\,+...
\label{eq412}
\end{equation}
\\
The $S^{(p)}$ are organized according to the power $p$ of the bosons 
$\beta,\beta^+, B, B^+ $. It is important to restate that, as in the case 
of the Hamiltonian expansion in eqs.~(\ref{eq320},\ref{eq321},\ref{eq322}),  
no normal ordering is assumed. Therefore a given term $S^{(p)}$ may very 
well contain $p-1, p-2,..$ powers of the bosons. The real expansion 
parameter in eq.~(\ref{eq320}) as well as in eq.~(\ref{eq412}) is $\sqrt{N}$.\\
The commutation relation eq.~(\ref{eq411}) will thus read
\\
\begin{equation}
 \left[ H\,,\, S \right] = \sum_{p} N^{3-\frac{p}{2}}\, C^{(p-1)} \,=\, 0.
\label{eq413}
\end{equation}
\\  
According to the basic observation that the polynomial is zero if 
and only if each of the monoms is vanishing, one can thus extract less 
stringent commutation relations between parts of the total Hamiltonian $H$ 
and the symmetry operator $S$. These are simply
\\
\begin{equation}
  C^{(p)} \,=\,  \left[ H^{(1)}\,,\, S^{(p+1)} \right]\,+\, 
  \left[ H^{(2)}\,,\, S^{(p)} \right]\,+\,.\,.\,.\,.\,+\,  
  \left[ H^{(p+1)}\,,\, S^{(1)} \right]\,=\, 0, \quad\quad\quad 
  \forall\, p\geq 0.
\label{eq414}
\end{equation}
\\
The relevant commutation relation to the order we are working is given 
by $C^{(1)}\,=\,0$ and reads explicitly
\\
\begin{equation}
   \left[ H^{(1)}\,,\, S^{(2)} \right]\,+\, \left[ H^{(2)}\,,\, 
   S^{(1)} \right] \,=\, 0 .
\label{eq415}
\end{equation}
\\
Recalling the fact that $H^{(1)}$ is eliminated by the construction of the 
variational (HB) basis, one is left only with 
\\
\begin{equation}
    \left[ H^{(2)}\,,\, S^{(1)} \right] \,=\, 0 .
\label{eq416}
\end{equation}
\\
This proves that a zero mode must exist. For the discussion below it
is useful to give an explicit form of the symmetry operator.
 Since $S^{(1)}$ contains only unit powers of the 
bosons $\beta$, $\beta^+$, $B$, $B^+$, the RPA operator
Eq. (50) is therefore the most general ansatz
with single boson excitations containing $S^{(1)}$. 
Explicitly, $S^{(1)}$ can be constructed as
\\
\begin{equation}
S^{(1)}\,=\,
\left[ U^{(1)}_{sp}({\vec 0})  \beta_0^+ \,-\,  V^{(1)}_{sp}({\vec 0}) 
\beta_0 \right] \quad+\quad 
\left[ U^{(2)}_{sp}({\vec 0},{\vec 0})  B^+_{00}
\,-\,  V^{(2)}_{sp}({\vec 0},{\vec 0}) B_{00} \right]
\label{eq59} 
\end{equation}
with
\begin{equation}
{\cal U}_{sp}({\vec 0},{\vec 0}) \,=\, -\,{\cal V}_{sp}({\vec 0},{\vec 0}) \,=\, \left(
\begin{array}{c} 
(2\pi)^{\frac{3}{2}} s \left(2\,{\cal E}_{\sigma}(0)\right)^{- \,\frac{1}{2}}
 {\cal E}_{\pi}(0) \\ -(2\pi)^3 \frac{s^2}{2} {\cal E}_{\pi}(0)
\end{array} 
\right).
\label{eq60} 
\end{equation}
\\
Physically this operator creates a zero-momentum sigma meson 
and a pion pair of vanishing total- as well as relative momentum.  

As mentioned above, it is instructive to express $H^{(2)}$ in the RPA 
basis. We shall follow the usual steps that can be found in the literature 
(see for instance \cite{RS80}). First it is easy to see that, by introducing
the RPA matrix, $H^{(2)}$ can be written in  matrix form as
\\
\begin{equation}
H^{(2)} \,=\, -\frac{1}{2} Tr\left({\cal A}-{\cal B}\right)\,\,+\,\, \frac{1}{2}\,\,
\left(\beta^+ \,\, \frac{1}{\sqrt{2}} B^+ \,\, \beta \,\,\frac{1}{\sqrt{2}} B \right)
 \left(
\begin{array}{cc}
{\cal A}\quad & \quad{\cal B} \\
\\ 
{\cal B}\quad & \quad{\cal A} 
\end{array} 
\right)
 \left(
\begin{array}{c}
\beta \\\frac{1}{\sqrt{2}}  B\\ \beta^+ \\ \frac{1}{\sqrt{2}}B^+
\end{array} 
\right).
\label{eq419}
\end{equation}
\\
Following \cite{RS80} we introduce a matrix $\Upsilon$ and denote the 
RPA matrix by ${\cal S}$:
\\
\begin{equation}
\Upsilon \,=\,  \left( \begin{array}{cccc}
{\cal U} & {\cal V}^{*}  \\
{\cal V} & {\cal U}^{*} 
\end{array} \right), \quad\quad\quad\quad
{\cal S} \,=\, \left(
\begin{array}{cc}
{\cal A} & {\cal B} \\
{\cal B} & {\cal A} 
\end{array} 
\right).
\label{eq420}
\end{equation}
\\
Then the closure relation of the RPA basis \cite{RS80} reads
\\
\begin{equation}
 \Upsilon {\cal N} \Upsilon^+ \, =\, \overline{\cal N}
\label{eq421}
\end{equation}
\\
where $\overline{\cal N}$ is a two by two norme matrice defined by substituting 1 to  $I_d$ in ${\cal N}$ of  eq.(\ref{eq332}). 
This  allows to  reexpress  the original boson operators 
in terms of RPA excitation operators as
\\
\begin{eqnarray}
 \beta^+_p &=&  \sum_{\nu >0}\left(
U^{(1)\, *}_{\nu}({\vec p})  Q^+_{\nu}(\vec p)  \,\, +\,\,      
V^{(1)}_{\nu}({\vec p})  Q_{\nu}(- \vec p)\right) \nonumber\\
 \beta_p &=&  \sum_{\nu >0}\left(
V^{(1)\, *}_{\nu}({\vec p})  Q^+_{\nu}(- \vec p)  \,\, +\,\,      
U^{(1)}_{\nu}({\vec p})  Q_{\nu}( \vec p)\right) \nonumber\\
 B^+_{q,p} &=&  \sqrt{2} \sum_{\nu >0}\left(
 U^{(2)\,*}_{\nu}({\vec q},{\vec p}) Q^+_{\nu}({\vec p} + {\vec q})
 \,\, +\,\
 V^{(2)}_{\nu}({-\vec q},{-\vec p}) Q_{\nu}(-{\vec p} - {\vec q})\right)
\nonumber\\
 B_{q,p} &=& \sqrt{2} \sum_{\nu >0}\left(
 V^{(2)\,*}_{\nu}(-{\vec q},-{\vec p}) Q^+_{\nu}(-{\vec p} - {\vec q})
 \,\, +\,\
 U^{(2)}_{\nu}({\vec q},{\vec p}) Q_{\nu}({\vec p} + {\vec q})\right).
\label{eq422}
\end{eqnarray}
\\
Now, using the orthonormalisation condition of the RPA basis which in 
compact notation reads
\\
\begin{equation}
 \Upsilon^+ \overline{\cal N} \Upsilon \, =\, {\cal N}
\label{eq423}
\end{equation}
one can extract, with the help of the RPA equation, the components of the 
eigenvectors 
${\cal U}_{\nu}({\vec p},{\vec q})$ and ${\cal V}_{\nu}({\vec p},{\vec q})$
as defined in eq.~(\ref{eq332}) and explicitly given in eq.~(\ref{eq335}). 
Finally, with the RPA excitation operators as defined in 
eq.~(\ref{eq328}) the Hamiltonian takes the following form  
\\
\begin{equation}
H^{(2)} \,=\, -\frac{1}{2} Tr\left({\cal A}-{\cal B} \right)\,\,+\,\, 
 \frac{1}{2}\,\,\left(Q^+ \,\, Q \right) \Upsilon^+ {\cal S} \Upsilon 
 \left(
\begin{array}{c}
 Q \\ Q^+ 
\end{array} 
\right).
\label{eq424}
\end{equation}
\\
In the exact chiral limit we have seen that a zero mode (with 
$\Omega_{sp} =0$) is present in the set of RPA solutions which requires  
special treatment when inverting the RPA. The procedure is standard and 
can be found in  textbooks (see for instance \cite{RS80}). Taking into 
account the zero mode which appears in the chiral limit, the completeness of 
the RPA basis is again recovered and the Hamiltonian can be expressed as 
follows 
\\
\begin{equation}
H^{(2)} \,=\, -\frac{1}{2} Tr\left({\cal A}-{\cal B}\right)\,\, +\,\,  \frac{1}{2}\sum_{{\vec p},
\,\nu >0}\Omega_{\nu}({\vec p})  \quad + \quad \sum_{{\vec p},\,\nu >0}
\Omega_{\nu}({\vec p})\, Q^+_{\nu}({\vec p})\, Q_{\nu}({\vec p}) \quad +
\quad
\frac{\left(S^{(1)}\right)^2}{2\, {\cal I}_0 V }. 
\label{eq425}
\end{equation}
\\
$S^{(1)}$ is the ${\cal O}(1)$ symmetry generator given in eq.~(\ref{eq412}) 
and $ {\cal I}_0$ is the moment of inertia per unit volume (with V the volume 
of the system) which can be calculated using the 
Valatin-Thouless equations \cite{TV62}:
\\
\begin{eqnarray}
\left[ H^{(2)}\,,\, C^{(1)} \right] &=& S^{(1)} \nonumber\\
\left[ S^{(1)}\,,\, C^{(1)} \right] &=&  {\cal I}_0
\label{eq428}
\end{eqnarray}
\\
The operator $C^{(1)}$  is an anti-Hermitian operator 
proportional to the canonical variable $T^{(1)}$, conjugate to the zero
mode induced by $S^{(1)}$ such that
\\
\begin{equation}
 C^{(1)}\, =\, i {\cal I}_0 T^{(1)} \quad\quad {\rm with} \quad
\left[ T^{(1)}\,,\, S^{(1)} \right] = i  
 \label{eq429}
\end{equation}
\\ 
The Valatin-Thouless equations  in eq.~(\ref{eq428}) are sufficient to 
determine both $T^{(1)}$ and the moment of Inertia  
$ {\cal I}_0$/ unit volume. An explicit calculation gives
\\
\begin{eqnarray}
 T^{(1)} &=&  \frac{i}{(2\pi)^3 2 s^2 m_{\pi}\,+\, 1 }\left[\left(B^+_{00}
 \,-\,B_{00}\right) \,\,-\,\,\frac{\sqrt{(2\pi)^3 2 {\cal E}_{\sigma}}}
 {(2\pi)^3 2 s\, m_{\pi}} \left(\beta^+_0\,-\,\beta_0\right)\right]
    \nonumber\\
 {\cal I}_0 &=& \frac{s^2}{4}
\label{eq430}
\end{eqnarray}
The expression for the moment of inertia suggests the very 
intuitive picture that the inertia of the vacuum against 
a chiral rotation is simply given by the amount of the spontaneous breaking 
via the condensate, $s$. 
Of course in general the volume $V$ in eq.~(\ref{eq425}) is infinite and thus 
the kinetic energy term of chiral rotation in (\ref{eq425}) is zero. However, 
in relativistic heavy ion collisions, situations may occur where there are 
small droplets of pionic matter with spontaneously broken chiral symmetry, 
surrounded by regions in the symmetry restored phase. In such cases the 
inertia is finite and we must keep the corresponding term.\\
Before closing this section, we wish to address briefly the question of the 
energy-weighted sum rule (EWSR). For any hermitian operator, $F$, this sum 
rule, to leading order in the $1/N$-expansion, is given by
\\
\begin{equation}
m_F^1\,=\, \sum_{\nu>0 }\Omega_{\nu}|\bra{\nu}F\ket{RPA}|^2\,=\,
\frac{1}{2} \bra{\psi}[ F, [ H^{(2)}, F]]\ket{\psi}
\label{eq431}
\end{equation}
\\ 
where $\ket{\psi}$ is the coherent state specified in eq.~(\ref{eq310}).
The proof goes along the standard arguments which can be found for 
instance in \cite{RS80, Yann}.\\ 
Among other things such sum rules can serve as a test for the correctness of 
numerical calculations. For example, a particular simple result is obtained 
with $F({\vec x}) = {\hat \sigma}({\vec x})$ for which the EWSR gives
$m_F^1(q) =1$, where $q$ is the three-momentum carried by the $\sigma$-field. 
Further sum rules with different excitation operators will be discussed in a 
forthcoming paper.

This concludes the formal discussion of the $1/N$ expansion
and the properties of the ensuing RPA problem. 
We will now turn to the construction of the $\pi\pi$ $T$-matrix.

\section{The $\pi\pi$-scattering equation}
A $\pi\pi$-scattering equation can be deduced from 
eqs.~(\ref{eq329}, \ref{eq330}) by eliminating the sigma subspace with the 
help of a Feshbach projection.
This procedure adds to the original $\pi\pi$-contact term of the Lagrangian  
an effective part which corresponds to s-channel sigma exchange. To leading
order in the $1/N$-expansion the projected RPA equation is given by
\\
\begin{equation}
\int d^3 {\vec q}_1\left(
\left[{\cal E}_{\pi}({\vec q}_1) \,+\,{\cal E}_{\pi}({\vec p}-{\vec q}_1)
\right]
\delta({\vec q}_1-{\vec q}_2)\,I_d
\, + \, 
\Lambda (\Omega_{\nu}, {\vec p},{\vec q}_1, {\vec q}_2)
\right)
\left(\begin{array}{c}
 U^{(2)}_{\nu}({\vec p},{\vec q}_1) \\  V^{(2)}_{\nu}({\vec p},{\vec q}_1)
\end{array} \right) 
\, =\, \Omega_{\nu}({\vec p}) 
\left(\begin{array}{c}
 U^{(2)}_{\nu}({\vec p},{\vec q}_2) \\ -\, V^{(2)}_{\nu}({\vec p},{\vec q}_2)
\end{array}\right).
\label{eq501}
\end{equation}
\\
Here $I_d$ is the two by two identity matrix and the interaction 
matrix $\Lambda$ is given by
\\
\begin{eqnarray}
\Lambda (\Omega_{\nu}, {\vec p},{\vec q}_1, {\vec q}_2)\,& =& \,
\lambda_0^2 \frac{\Omega_{\nu}^2\,-\,{\cal E}_{\pi}^2({\vec p})}
{\Omega_{\nu}^2\,-\,{\cal E}_{\sigma}^2({\vec p})}
 \left( \begin{array}{cc} 1 & 1 \\ 1  & 1 \end{array} \right)
\frac{d ({\vec p},{\vec q}_1, {\vec q }_2) }
{(2\pi)^3}\nonumber\\
d ({\vec p},{\vec q}_1, {\vec q }_2) \,& =&\, \left[ 16 
{\cal E}_{\pi}({\vec q}_1){\cal E}_{\pi}({\vec q}_2)
{\cal E}_{\pi}({\vec p}-{\vec q}_1){\cal E}_{\pi}({\vec p}-
{\vec q}_2)\right]^{-1/2}
\label{eq502}
\end{eqnarray}
\\
From eqs.~(\ref{eq501}) and (\ref{eq502}) it is now straightforward to 
construct the corresponding two-pion propagator in the scalar-isoscalar 
channel:
\\
\begin{equation}
G_{\pi\pi}(t-t', {\vec x}_1,{\vec x}_2,{\vec x}_3,{\vec x}_4)\,\,=\,\,
-i\, \bra{RPA}T \left( {\vec \pi}({\vec x}_1) {\vec \pi}({\vec x}_2)\right)_t
 \left( {\vec \pi}({\vec x}_4) {\vec \pi}({\vec x}_3)\right)_{t'} \ket{RPA}
\label{eq503}
\end{equation}
\\
where $T$ denotes the time-ordering operator and $\ket{RPA}$ is the 
RPA-correlated vacuum. The Green's function only depends on the time 
difference $t-t'$ but, in spite of being non-covariant, still contains the 
full dynamics at this order since t or u-channels only appear in next 
order. Taking the Fourier transform of $G_{\pi\pi}$ and keeping 
the relevant part to leading order of the expansion in the new bosons 
$B, B^+$  gives
\\
\begin{equation}
G_{\pi\pi}(t-t', {\vec q}_1 ,{\vec q}_2 ,{\vec q}_3 ,{\vec q}_4) = N\,
d ({\vec q}_1,{\vec q}_2, {\vec q }_3, {\vec q }_4) 
\bra{RPA}T  \left(B_{q_1,q_2 }\,+\,B^+_{-q_1,-q_2}\right)_t
\left(B_{q_3,q_4}\,+\,B^+_{-q_3,-q_4}\right)_{t'}\ket{RPA}
\label{eq504}
\end{equation}
\\
The Fourier transform of the time difference $t-t'$ defines the 
center of mass energy $E$ . At this order the two-pion Green's function 
is related in the usual way to the $T$-matrix
\\
\begin{equation}
G_{\pi\pi}(E, {\vec q}_1 ,{\vec q}_2 ,{\vec q}_3 ,{\vec q}_4) = 
G_{\pi\pi}^0(E, {\vec q}_1 ,{\vec q}_2 ,{\vec q}_3 ,{\vec q}_4) +
 G_{\pi\pi}^0(E, {\vec q}_1 ,{\vec q}_2 )
 T(E, {\vec p})
G_{\pi\pi}^0(E, {\vec q}_3 ,{\vec q}_4) 
\label{eq804}
\end{equation}
and a  straightforward reduction then gives the $T$-matrix as
\\
\begin{eqnarray}
 \int d{\vec k}_1d{\vec k}_2d{\vec k}_3d{\vec k}_4\,\, 
G^{0\, -1}_{\pi\pi}(E, {\vec q}_1 ,{\vec q}_2 ,{\vec k}_1 ,{\vec k}_2)
&G_{\pi\pi}(E, {\vec q}_1 ,{\vec q}_2 ,{\vec q}_3 ,{\vec q}_4)\,\,
G^{0\, -1}_{\pi\pi}(E, {\vec k}_3 ,{\vec k}_4 ,{\vec q}_3 ,{\vec q}_4)
\nonumber\\
 =&\,\,\frac{1}{(2\pi)^3}\, 
 \delta({\vec q}_1+{\vec q}_2-{\vec q}_3-{\vec q}_4)\,  
T_{\pi\pi}( E, {\vec p})
\label{eq505}
\end{eqnarray}
\\
where ${\vec p}$ is the total three-momentum of the pion pair, 
$ {\vec p} ={\vec q}_1+{\vec q}_2 ={\vec q}_3+{\vec q}_4$. On the other 
hand $G^{0}_{\pi\pi}$ is the free two-pion Green's function defined as
\\
\begin{eqnarray}
G^0_{\pi\pi}(t-t', {\vec q}_1 ,{\vec q}_2 ,{\vec q}_3 ,{\vec q}_4) &=& N\,
d ({\vec q}_1,{\vec q}_2, {\vec q }_3, {\vec q }_4) 
\bra{\psi}T  \left(B_{q_1,q_2 }\,+\,B^+_{-q_1,-q_2}\right)_t
\left(B_{q_3,q_4}\,+\,B^+_{-q_3,-q_4}\right)_{t'}\ket{\psi}\nonumber\\
\nonumber\\
&=& G^0_{\pi\pi}(t-t',{\vec q}_1,{\vec q}_2)\left[\delta({\vec q}_1-{\vec q}_3)\delta({\vec q}_2-{\vec q}_4)\, +\,
\delta({\vec q}_1-{\vec q}_4)\delta({\vec q}_2-{\vec q}_3)\right]  
\label{eq506}
\end{eqnarray}
\\
where $\ket{\psi}$ is the coherent state defined earlier. In energy space 
one has
\\
\begin{equation}
G^0_{\pi\pi}(E, {\vec q}_1 ,{\vec q}_2 ) =
\frac{{\cal E}_{\pi}({\vec q}_1)\,+\,{\cal E}_{\pi}({\vec q}_2) }
{2\, {\cal E}_{\pi}({\vec q}_1){\cal E}_{\pi}({\vec q}_2)} \,\,
\frac{1} {E^2\,-\,\left({\cal E}_{\pi}({\vec q}_1)\,+\,
{\cal E}_{\pi}({\vec q}_2)\right)^2\, +\, i \eta}
\label{eq507}
\end{equation}
\\
Using the Heisenberg picture, the reduction in eq.~(\ref{eq505}) can be 
performed explicitly  with the help of the RPA representation in 
eqs.~(\ref{eq422}, \ref{eq425}) as well as the RPA solutions in 
eqs.~(\ref{eq333}, \ref{eq334}, \ref{eq335} , \ref{eq336}). This yields the 
following spectral representation for the T-matrix 
\\
\begin{equation}
T_{\pi\pi}(E, {\vec p}) \,=\, \frac{1}{2} \sum_{\nu > 0} \left|\frac{V_{\pi\pi}(\Omega_{\nu}, {\vec p})}{\Xi(\Omega_{\nu},{\vec p})}\right|^2 \left( \frac{1}{E\,-\,\Omega_{\nu}({\vec p}) \,+\, i\eta}\,-\, \frac{1}{E\,+\,\Omega_{\nu}({\vec p}) \,-\, i\eta}\right)  
\label{eq508}
\end{equation}
\\
where the  $V_{\pi\pi}$ and $\Xi$ functions have been given in 
eq.~(\ref{eq336}).
An equivalent way of solving the scattering problem in the RPA is to extract 
the bare $\pi\pi$-interaction $V_{\pi\pi}$
which serves as the kernel of a Lippmann-Schwinger equation:
\\
\begin{equation}
T_{\pi\pi}(E, {\vec p}) \,=\, V_{\pi\pi}(E, {\vec p})\, +\, \frac{1}{2}\int 
\frac{d{\vec q}}{(2\pi)^3}\, V_{\pi\pi}(E, {\vec p}) G_{\pi\pi}^0(E, {\vec p}, 
{\vec q}) T_{\pi\pi}(E, {\vec p})
\label{eq509}
\end{equation}
\\
with $G_{\pi\pi}^0$ as given in ~(\ref{eq507}) and the bare interaction 
$V_{\pi\pi}$ obtained by reducing the Green's function in eq.~(\ref{eq503}) 
to lowest order in the dimensionless coupling constant, $\lambda_0^2$. 
The solution of the above equation has a simple algebraic form 
\\
\begin{eqnarray}
T(E, {\vec p}) \,&=& \frac{V_{\pi\pi}(E, {\vec p})}
{1\,\,-\,\, \frac{1}{2}V_{\pi\pi}(E, {\vec p})\Sigma_{\pi\pi}(E, {\vec p})}
\nonumber\\
\nonumber\\
V_{\pi\pi}(E, {\vec p})\,&=&\, 
\,2\lambda^2_0 \frac{ E^2\,-\,{\cal E}_{\pi}^2({\vec p})}
{E^2 \, -\, {\cal E}_{\sigma}^2({\vec p})} \nonumber\\
\nonumber\\
\lambda_0^2 \,&=&\, N \frac{{\cal E}_{\sigma}^2({\vec 0}) - 
{\cal E}^2_{\pi}({\vec 0})}{2 f_{\pi}^2}\quad\quad\quad with\quad f_{\pi}^2 = 
N s^2 
\label{eq510}
\end{eqnarray}
\\ 
where $\Sigma_{\pi\pi}(E, {\vec p})$ denotes the $\pi\pi$-bubble defined in 
eq.~(\ref{eq334}). The two forms of the T-matrix given above are equivalent. 
This can be verified by taking the residues at each pole of 
$T_{\pi\pi}$ from the expression in eq.~(\ref{eq510}) which then directly leads to the spectral 
representation in eq.~(\ref{eq508}).
From eqs.~(\ref{eq504}, \ref{eq804}, \ref{eq505}, \ref{eq510}) and from the 
imaginary part of the total two-pion propagator $Im G_{\pi\pi}$
one deduces that the spectrum starts at $2m_{\pi}$ i.e. it is gapless in the 
chiral limit, in agreement with the Goldstone Theorem (see sect. 4).\\   
The $\sigma$ -propagator can also be calculated starting from the related 
two-point Green's function
\\
\begin{equation}
D_{\sigma}(t_1-t_2, {\vec x}_1,{\vec x}_2)  \,\,=\,\,
-i\, \bra{RPA}T  {\sigma}({\vec x}_1,t_1) {\sigma}({\vec x}_2, t_2)\ket{RPA}
\label{eq511}
\end{equation}
\\
which, after Fourier transformation, can be written in  spectral  
representation as
\\
\begin{equation}
D_{\sigma}(E, {\vec p}) \,=\, \frac{1}{2}\sum_{\nu > 0} \left|\frac{V_{\pi\pi\sigma}}
{\Xi(\Omega_{\nu},{\vec p})}\frac{1}{\Omega_{\nu}^2 - {\cal E}_{\sigma}^2
({\vec p})}\right|^2 \left( \frac{1}{E\,-\,\Omega_{\nu}({\vec p}) \,+\, 
i\eta}\,-\, \frac{1}{E\,+\,\Omega_{\nu}({\vec p}) \,-\, i\eta}\right).  
\label{eq512}
\end{equation}
\\
Alternatively one can obtain $D_\sigma$  by solving a Dyson 
equation with the mass operator containing the $\pi\pi$-RPA fluctuations.
This leads to 
\\
\begin{equation}
D_{\sigma}(E, {\vec p}) \,=\, \left[{ E^2\,-\, {\cal E}_{\sigma}^2
({\vec p})\,-\, \frac{2 \lambda_0^4 s^2\,{\Sigma}_{\pi\pi}(E,{\vec p})}
{ 1\,-\,  \lambda_0^2 {\Sigma}_{\pi\pi}(E,{\vec p})}}\right]^{-1}
 \label{eq513}
\end{equation}
\\
and again these two forms of $D_\sigma$ are equivalent,
as can be seen by taking the residue at each pole in
eq.~(\ref{eq513}) which yields the spectral form (\ref{eq512}). From these 
observations the scattering matrix $T_{\pi\pi}$ can be reexpressed as
\\
\begin{equation}
T_{\pi\pi}(p^2) \,=\, \frac{ D_{\pi}^{-1}(p^2)\,-\,D_{\sigma}^{-1}(p^2)}
{s^2} 
\frac{ D_{\sigma}(p^2)}{D_{\pi}(p^2)}
\label{eq514}
\end{equation}
\\
where $p^2 = E^2 - {\vec p}^2$. 
As we have seen above to this order in the $1/N$ expansion the pion is 
obtained in the Hartree approximation, and $D_{\pi}$ is therefore given by
\\
\begin{equation}
 D_{\pi}(p^2) \,=\, \frac{1}{p^2 \,-\, m_{\pi}^2}.
\label{eq515}
\end{equation}
\\
Thus one pion can be taken soft ({\sl i.e.} $m_\pi=0$) in which case the above
expression reduces to a Ward identity, linking the pion four-point function 
to the sigma and pion two-point functions \cite{BLEE}. The expression in 
(\ref{eq514}) is more general, however, and includes the soft-pion limit as a 
special case. Finally it can be seen that, at the physical threshold 
$ E^2 = 4 m_{\pi}^2, {\vec p} ={\vec 0}$, the T-matrix  
is directly proportional to the pion mass and hence vanishes in the chiral 
limit. This is, of course, required by the Goldstone nature of the pions.  
Away from the chiral limit it is interesting to compare the $a_0^0$ 
scattering length resulting from the T-matrix in eq.~(\ref{eq510}) with
the well-known tree level results. To lowest order in the coupling constant 
and for $N = 3$, the $a_0^0$ scattering length can be read off 
from eq.~(\ref{eq510}) and yields
\\
\begin{equation}
a_{0\, tree}^0\,=\frac{9}{32 \pi}\frac{ m_{\pi}}{f_{\pi}^2}
\label{eq516}
\end{equation}
\\
while the standard tree-level value is 
$a_0^0 = \frac{7}{32 \pi}\frac{ m_{\pi}}{f_{\pi}^2} $. The difference 
stems from the fact that, to leading-order in $1/N$,  only   
s-channel contributions appear while the standard Weinberg tree-level 
result contains $t$- and $u$-channel pieces as well.  
These contributions are known to be repulsive and, in the $1/N$ expansion,
they only enter in next-to-leading order. It is tempting to obtain
the $a^0_0$ scattering length for the full T-matrix. Unfortunately
for any reasonable choice of parameters (the cut-off, $\Lambda$, and the 
physical sigma mass, ${\cal E}_\sigma$) no acceptable fit of the 
scalar-isoscalar phase shifts can be obtained from eq.~(\ref{eq510}). 
This is to be expected since the RPA resummation of the scattering 
series will increase $a_0^0$ further from its tree level value, which 
is already too large. This hints to the fact that higher orders in the $1/N$
expansion will be crucial to get a quantitative description of the 
empirical $s-$wave phase shifts. It will be very interesting to work these out 
explicitly, exhibiting u- and t- channel contributions. This will be done in 
the future.

\section{Summary and Outlook}
In an attempt to obtain a symmetry-conserving $\pi\pi$-scattering equation,
consistent with unitarity, we have investigated the cut-off version of the 
$O(N+1)$ linear sigma-model. Increasing interest in connection with two-pion 
correlations in hadronic matter makes such studies particularly timely. On the 
one hand, as is well known, two-pion correlations play a crucial role in the 
N-N interaction and studies of the in-medium corrections are still in their 
infancy \cite{ARCSW,RDW}. On the other hand, in purely pionic matter, 
the $\pi\pi$-interaction, due 
to Bose statistics, may become strongly enhanced such that, in the 
scalar-isoscalar channel, two-pion bound states are likely to appear at higher
temperature. Obviously, such features cannot be treated in chiral perturbation 
theory and non-perturbative schemes, respecting constraints from chiral 
symmetry, are called for. The present paper provides a novel approach 
by applying well-known Boson-expansion techniques. That such techniques,
which are highly developed in nuclear physics, can be very useful
for interacting Bose systems  was first demonstrated by (CDDF-BPD) 
\cite{CDDF, PIDUK}. We have applied this approach, for the first time, in the context of field theory. In a systematic $1/N$-expansion it has become possible 
to derive a $\pi\pi$-scattering equation, to lowest order in $1/N$.
This scattering equation is of RPA type and fulfills all required properties 
such as unitarity, Ward identities and standard RPA sum rules. The formalism 
is  manifestly non-covariant. However the full covariant theory collapses to a two times-theory in this order of the $1/N$ expansion. Therefore no
breaking of covariance occurs in the final results. 
 Furthermore special care 
had to be taken in dealing with Goldstone modes, which is crucial for 
$\pi\pi$-scattering in the chiral limit. Due to the symmetry-conserving
character of the approximations these aspects are treated  correctly in the present approach.
The lowest-order equation disregards u- and t-channel exchanges in the 
$I=J=0$ channel and no physically reasonable parameters 
(cut-off and sigma mass) could be found to reproduce the empirical 
$\pi\pi$ phase shifts. Contributions from u- and t-channel exchanges ,
possibly curing this problem, enter only in next-to-leading order and will 
be studied in a forthcoming publication. 

One can envision several other applications of interest. A direct extension
of the present study would be the inclusion of fermionic degrees of freedom to 
treat the $\pi-N$ system on a new level, including non-perturbative $\pi\pi$ 
rescattering effects systematically. 
Another area concerns the thermodynamics of the linear sigma-model. 
In ref.~\cite{ACSW} the finite-temperature 
version of the HFB-QRPA in $O(4)$ has been studied, in an attempt to describe 
the chiral phase transition. This transition was found to be of first-order. To
leading order in the $1/N$-expansion, on the other hand, it turns out to be of 
second-order, as will discussed in future work. Similar to
the $\pi\pi$-scattering case it will be important to see what the 
next-to-leading-order corrections are.
\vspace*{0.3cm}

\underline {\bf Acknowledgments:} We would like to thank G. Chanfray 
for discussions and for his interest in this work. We also thank H.P. Pavel
for a careful reading of the manuscript.  
This work was supported in part by a grant from the National Science
Foundation, NSF-PHY-94-21309. One of us (Z. A.) acknowledges financial
 support from  Gesellschaft f\"{u}r Schwerionenforschung (GSI-Darmstadt). 

\newpage
\section{Appendix 1}
In this appendix we want to show how one can handle safely  the square 
roots originating from the Holstein-Primakoff mapping. Obviously one rather 
needs  the expanded form of these functions in terms of the operators
$A$ and $A^+$. Thus one could think of performing a formal Taylor expansion.
However since these operators are non commuting one  therefore 
must be careful when proceeding with the expansion. The major difficulty comes from the operator differentiation needed for the evaluation of each order in the Taylor expansion. Several methods have been designed for this purpose. Among these methods there exists a very simple and  powerful one based on the parameter differentiation technique (see for instance \cite{AIZU}). Here we will present a slightly different derivation.\\
The starting point is the parameter differentiation of the exponential. Given a function $z$ of the complex variable $\lambda$ then the following identity
holds: 
\\
\begin{equation}
\frac{\partial e^{z(\lambda)}}{\partial \lambda} \,=\, \int_{0}^{1} 
d\alpha\,\, e^{(1-\alpha)z}\, \frac{\partial z(\lambda)}{\partial \lambda}
\, e^{\alpha z}
\label{eq600}
\end{equation}
\\
The identity above was derived by Snider in \cite{SNID} and can be recovered, following the same author, by first recalling the relation:
\\
\begin{equation}
\frac{\partial z^n}{\partial \lambda} \,=\, \sum_{m=0}^{n-1}\, z^{m} 
\, \frac{\partial z}{\partial \lambda}
\,  z^{n-m-1}
\label{eq601}
\end{equation}
\\    
which can be shown to hold by a simple induction. Using this result the left
hand side of eq.(\ref{eq600}) reads:
\\
\begin{equation}
\frac{\partial e^{z(\lambda)}}{\partial \lambda} \,=\, \sum_{n=0}^{\infty}\,
\sum_{m=0}^{n-1}\, \frac{1}{n!} 
 z^{m} \, \frac{\partial z}{\partial \lambda}\,  z^{n-m-1}
\label{eq602}
\end{equation}
\\
Making now the substitution: $ p\,=\,n-m-1 $,  one observes that 
\\
\begin{equation}
\frac{m!\,p!}{(p+m+1)!} \,=\, B(p+1\, ,\, m+1) \,=\, \int_{0}^{1}  \left(1-\alpha\right)^m \alpha^p\,d\alpha
\label{eq603}
\end{equation}
\\
where $B(x,y)$ is nothing else but the  beta function. Inserting now the above in eq.(\ref{eq602}) one gets:
\\
\begin{eqnarray}
\frac{\partial e^{z(\lambda)}}{\partial \lambda} \,&=&\, 
\sum_{m=0}^{\infty}\,\sum_{p=0}^{\infty}\,\, \frac{m! p!}{(p+m+1)!} 
 \frac{z^{m}}{m!} \,\, \frac{\partial z}{\partial \lambda}\,\,
\frac{z^{p}}{p!} \nonumber\\
 \,&=&\, \int_{0}^{1}d\alpha\, \sum_{n,p =0}^{\infty} \,\,
 \left(1-\alpha \right)^m \, \alpha^p 
\frac{z^{m}}{m!} \,\, \frac{\partial z}{\partial \lambda}\,\,
  \frac{z^{p}}{p!} 
\label{eq604}
\end{eqnarray}
\\
Now one can easily group the terms together and check that this is effectively the desired result of eq.(\ref{eq600}).\\ 
Back to our problem of the square root expansion, we wish to make use of the identity derived so far. First to generate an exponential out of the square root one simply recalls the inverse Laplace transform:
\\
\begin{equation}
\sqrt{t} = \frac{1}{4\sqrt{\pi}i}\int_{c-i\infty}^{c+i\infty} e^{ts} s^{-\frac{3}{2}} ds
\label{eq605}
\end{equation}
\\  
Expanding the Holstein-Primakoff square root amounts then to expanding the exponential of the same argument times an arbitrary parameter $s$. For the purpose of the parameter differentiation we define the argument of the square root (
or of the exponential) as:
\\
\begin{equation}
t  \,=\,  t(\lambda) \,=\, N\,\, +\,\, {\hat d}^2 \,\,+\,\, \lambda \left({\hat d} {\tilde A} +  {\tilde A}^+{\hat d} +{\tilde A}^+{\tilde A}\right) 
\label{eq606}
\end{equation}
\\  
For $\lambda = 1$ one recovers the argument of the HP square root. We recall that the matrix ${\hat d}$ is diagonal,and that it does not commute with the operators ${\tilde A}$ and ${\tilde A}^+$. The fact that it scales like $\sqrt{N}$ is crucial because it helps to organize the expansion in the terms of powers of $\sqrt{N}$ which is the real parameter of the expansion (see text).\\
Gathering everything together one can write finally the HP square root into the following Taylor expansion:
\\
\begin{equation}
\left(\sqrt{t(\lambda)}\right)_{q,p} = \frac{1}{4\sqrt{\pi}i}\int_{c-i\infty}^{c+i\infty} s^{-\frac{3}{2}} ds \left( 
\left[e^{s t(\lambda)}\right]_{\lambda =0}\,\,+\,\, \lambda \left[\frac{\partial e^{s t(\lambda)}}{\partial \lambda}\right]_{\lambda =0} \,\,+\,\,
\frac{\lambda^2}{2}  \left[\frac{\partial^2 e^{s t(\lambda)}}{\partial \lambda^2}\right]_{\lambda =0}\,+\,
 \ldots\ \right)_{q,p}
\label{eq607}
\end{equation}
\\ 
As an application of the above considerations one can, through a trivial intermediate algebra, show that the lowest orders (used in this paper) for the
HP square root read:
\\
\begin{eqnarray}
 \left(A^+\sqrt{ N\,+\,A^+A}\right)_{q,p}\,&=& N \left[d_q\sqrt{1+d_q^2} \delta_{q, p}\right] + N^{\frac{1}{2}} \left[{\tilde A}^+_{q,p}\sqrt{1+d_p^2}\,+\,d_q \frac{\sqrt{1+d_q^2}-\sqrt{1+d_p^2}}{d_q^2-d_p^2}\left(d{\tilde A}+{\tilde A}^+d\right)_{q,p}\right]\nonumber\\
&+&\, \left[ \sum_{m} \frac{\sqrt{1+d_m^2}-\sqrt{1+d_p^2}}{d_m^2-d_p^2}
\left(\delta_{q,m}\, d_q \left({\tilde A}^+{\tilde A}\right)_{q,p} \,+\, {\tilde A}^+_{q,m}\left(d{\tilde A}+{\tilde A}^+d\right)_{m,p}\right)\right. \nonumber\\
&+&\quad \left. \sum_{m} {\cal F}(q,m,p)  \frac{d_q}{2} \left(d{\tilde A}+{\tilde A}^+d\right)_{q,m} \left(d{\tilde A}+{\tilde A}^+d\right)_{m,p}
\right] \,\,+\quad\quad \theta(N^{-\frac{1}{2}}) \ldots\
\label{eq608}
\end{eqnarray}
\\
with the following definition of the function ${\cal F}(q,m,p)$:
\begin{equation}
{\cal F}(q,m,p)=  
 \frac{\sqrt{1+d_q^2}+\sqrt{1+d_p^2}- 2\sqrt{1+d_m^2}}{(d_q^2-d_m^2)
(d_m^2-d_p^2)} \,+\,
 \frac{\sqrt{1+d_q^2}-\sqrt{1+d_p^2}}{d_q^2-d_p^2}\left(\frac{1}{(d_q^2-d_m^2)}+\frac{1}{(d_p^2-d_m^2)}\right)
\label{eq609}
\end{equation}
\\
One can further simplify these expressions by using the  symmetry property  of the bosons ${\tilde A}, {\tilde A}^+$ under the interchange of the indices $q$ and $p$. The same expansion can be performed for the hermitian conjugate of the right hand side of eq.(\ref{eq608}). These two expansions are the building blocks of the Hamiltonian in eq.(\ref{eq306}).


\end{document}